\documentclass[aps,superscriptaddress,twocolumn,a4paper,10pt]{revtex4}
\usepackage[colorlinks=true, pdfstartview=FitV, linkcolor=blue, citecolor=red, urlcolor=black]{hyperref}
\usepackage{graphicx}
\usepackage[all]{xy}
\usepackage{amsmath}
\usepackage{amssymb}
\usepackage{tensor}
\usepackage{orcidlink}
\usepackage{amsthm}
\usepackage{comment}
\newcommand{\be}{\begin{equation}}
\newcommand{\ee}{\end{equation}}
\newcommand{\ben}{\begin{eqnarray}}
\newcommand{\een}{\end{eqnarray}}
\newcommand{\bes}{\begin{subequations}}
\newcommand{\ees}{\end{subequations}}
\def\bal#1\eal{\begin{align}#1\end{align}}

\newcommand{\bfi}{\begin{figure}}
\newcommand{\efi}{\end{figure}}
\newcommand{\bc}{\begin{center}}
\newcommand{\ec}{\end{center}}

\newcommand{\sech}{{\rm sech}}

\newcommand{\arcsinh}{{\rm arcsinh}}

\newcommand{\LL}{{\cal L}}

\newcommand{\Sc}{{\cal S}}

\begin{document}

\title{Compact structures in impurity-doped vacuumless systems}
\author{I. Andrade\,\orcidlink{0000-0002-9790-684X}}
        \email[]{andradesigor0@gmail.com}\affiliation{Departamento de F\'\i sica, Universidade Federal da Para\'\i ba, 58051-970 Jo\~ao Pessoa, PB, Brazil}
\author{D. Bazeia\,\orcidlink{0000-0003-1335-3705}}\email[]{ bazeia@fisica.ufpb.br}\affiliation{Departamento de F\'\i sica, Universidade Federal da Para\'\i ba, 58051-970 Jo\~ao Pessoa, PB, Brazil}
\author{M.A. Marques\,\orcidlink{0000-0001-7022-5502}}
        \email[]{marques@cbiotec.ufpb.br}\affiliation{Departamento de Biotecnologia, Universidade Federal da Para\'\i ba, 58051-900 Jo\~ao Pessoa, PB, Brazil}
\author{R. Menezes\,\orcidlink{0000-0002-9586-4308}}
     \email[]{rmenezes@dcx.ufpb.br}\affiliation{Departamento de Ci\^encias Exatas, Universidade Federal
da Para\'{\i}ba, 58297-000 Rio Tinto, PB, Brazil}

\begin{abstract}
We investigate novel structures which arise from the compactification of vacuumless kinks in scalar field models coupled to impurities that preserve half the BPS sectors, described by first-order equations. We also investigate the behavior of the energy density and linear stability of the solutions. We show that compact vacuumless kinks cannot be obtained in impurity-free canonical models. By considering two distinct impurities, we study the conditions needed to induce compactification. In this scenario, stable half-compact or compact solutions are shown to emerge from the systems.  
\end{abstract} 

\maketitle

\section{Introduction}
Topological structures arise in Field Theory under the action of scalar and other fields as static configurations which connect the vacua of the systems \cite{manton}. The canonical models capable of supporting such objects consist of the difference between dynamical and potential terms. Among the aforementioned structures, the simplest ones are kinks, which appear in a single spatial dimension and are compatible with a first-order framework based on the minimization of the energy \cite{bogo}. A well-known potential which supports them is the so-called $\phi^4$, which engenders two degenerate minima connected by a field of a hyperbolic tangent profile that falls off exponentially.

In canonical models, the fall off of the solutions depends on the minima of the potential. For instance, the aforementioned exponential tails appear for potentials with finite second derivative at the minima. In the literature, several types of structures with distinct asymptotic behavior can be found \cite{long1,long2,duploexp,expexpexp,superlg1,superlg2,vacuumless1,vacuumless2,comp1,comp2}. In particular, an interesting class of models was proposed in Refs.~\cite{vacuumless1,vacuumless2}, whose potentials do not support minima but vanish asymptotically, so they are usually called \emph{vacuumless} or \emph{runaway}; see also \cite{vacuumless3,vacuumless4}. They give rise to stable field configurations which range from $-\infty$ to $\infty$ (or from $\infty$ to $-\infty$) as the spatial coordinate goes from $-\infty$ to $\infty$. Although such solutions have logarithmic divergence, the energy density is localized and the total energy is finite. It is also possible to obtain semi-vacuumless models, in which only one of the tails of the solutions diverge  \cite{semivl1}. Vacuumless systems have been used in many contexts over the years, such as in gravitation and cosmology \cite{vlgrav1,vlgrav2,vlgrav3,vlcosmo}, tachyonic dynamics \cite{vltac1,asen1,asen2,trilogia1}, collisions \cite{vlcoli} and vortices \cite{vlvortex}.

Another type of structures which emerge in canonical models are the compact ones. They firstly appeared in the presence of V-shaped potentials \cite{comp2,comp23,comp24}. Later, it was shown that they can be obtained by introducing a parameter which is able to compactify the solution by inducing an infinite second derivative at the minima of the potential \cite{k2c}. Compact solutions reach the minima of the potential at a finite location, whose associated interval define the width of the solutions. Interestingly, the energy density is fully concentrated inside the compact space, vanishing otherwise. Of course, there is also the possibility of compactifying only one of the tails, leading to half-compact configurations \cite{comp7}. Over the years, several papers have appeared dealing with localized structures that generate compact profile \cite{kdvcomp1,kdvcomp2,kdvcomp3,comp222,comp3,comp4,comp45,comp5,comp6,comp8,comp89,comp899,comp8999,comp9,comp10,compschro}.

In the last century, between the 70s and 90s, several works began to consider the inclusion of impurities in the canonical models, which take into account the presence of spatial inhomogeneities \cite{imp1,imp2,imp3,imp4,imp5,imp6,imp7,imp8}. Recently, important developments were made in this direction, related to the addition of impurities which preserve half the BPS sectors \cite{imp9,imp10,imp11,imp12,PRL,PRD,imp13,SW1,SW2,imp14,colisaolongimp,impureza5,impureza6} in $(1,1)$-dimensional models. In this scenario, one can find minimum energy configurations described by first-order equations in which the impurity is additive, modifying the slope of the solutions. This leads to novel results, such as stable non-monotonic kinks. Moreover, the aforementioned first-order formalism satisfies Derrick's rescaling argument \cite{derrick}, which is related to stability under contractions and dilations, via the stressless condition \cite{impureza5,impureza6}. The study of scalar field models with impurities has gained attention in the latter years, particularly in the context of collisions, where spectral walls may arise \cite{PRL,PRD,imp13,SW1,SW2,imp14,colisaolongimp,forcelongimp}.

In this work, we show how to induce the compactification of the vacuumless solutions found in \cite{vacuumless1,vacuumless2} via the addition of impurities. In Sec.~\ref{secmodel}, we present the scalar field model coupled to impurities, showing the equations of motion, first-order formalism, energy and linear stability. We also make a brief review about the impurity-free vacuumless solutions and show that they cannot be compactified in the canonical model without impurities. In Sec.~\ref{secexamples}, we investigate two distinct impurities, both containing a parameter $\alpha$ which is able to compactify the solutions as it gets larger and larger, giving rise to vacuumless solutions with half-compact or compact support. Interestingly, depending on the initial condition used to solve the first-order equation, singular kinks similar to the one found in \cite{asen1,asen2} may arise. We finish the work in Sec.~\ref{secoutlook}, where we summarize our findings and add new comments on perspectives for future research.

\section{Scalar field models with impurities}\label{secmodel}
Let us consider the model proposed in Refs.~\cite{imp9,imp10} in $(1,1)$ spacetime dimensions, with metric tensor $\eta_{\mu\nu} = \text{diag}(1,-1)$ and the scalar field $\phi$ coupled to the impurity $\sigma(x)$ via de action $\Sc = \int dx\, dt\,\LL$, where the Lagrangian density is
\be\label{lmodel}
\LL = \frac12\dot{\phi}^2-\frac12\left(\phi'-\sigma(x)\right)^2-V(\phi)-\sigma(x)G(\phi).
\ee
In the above expression, the dot and the prime represent derivatives with respect to time and space, respectively. Also, $V(\phi)$ is the potential and $\sigma(x)$ is not an auxiliary field. Actually, $\sigma(x)$ is a background field which, to our purpose, is static. Notice that the impurity function not only is coupled via the derivative term, but also to the self-interaction of $\phi$ via $G(\phi)$. The equation of motion for time-dependent field, $\phi(x,t)$, is
\be\label{eom}
\Ddot{\phi} - \phi''+\sigma' +V_\phi+\sigma(x)G_\phi = 0.
\ee
The energy density of the field configurations can be calculated standardly, via $\rho=\dot{\phi}^2-\LL$, which reads
\be\label{rho}
\rho =\frac12\dot{\phi}^2+ \frac12\left(\phi'-\sigma(x)\right)^2+V(\phi)+\sigma(x)G(\phi).
\ee
By integrating it, we get the energy $E=\int_{-\infty}^\infty dx\,\rho$. By introducing an auxiliary function $W=W(\phi)$, the above expression can be written in the form 
\be
\begin{aligned}
E &= \int_{-\infty}^\infty dx\,\bigg(\frac12\dot{\phi}^2+\frac12\left(\phi'-\sigma(x)-W_\phi\right)^2\\
    &+V(\phi)-\frac12W_\phi^2+\sigma(x)(G(\phi)-W_\phi)+W'\bigg).
\end{aligned}
\ee
If the functions which control the self-interactions of the field are given by
\be\label{VGW}
V(\phi) = \frac12W_\phi^2 \quad\text{and}\quad G(\phi)=W_\phi,
\ee
the energy is bounded, $E\geq E_B$, where
\be
E_B = W(\phi(\infty))- W(\phi(-\infty))
\ee
is the minimum value. The bound is saturated if the solutions are static, $\dot\phi=0$, and first-order equation
\be\label{fo}
\phi'=W_\phi + \sigma(x)
\ee
is satisfied. This expression is compatible with the stressless condition, which is required to ensure the stability under contractions and dilations \cite{impureza5,impureza6}; this equation also leads to stability against spatial translations. By using Eq.~\eqref{fo}, we can write the energy density as
\be\label{rhow}
\rho(x) = (W_\phi+\sigma(x))W_\phi.
\ee

We now verify the linear stability of the solutions, by substituting  $\phi(x,t)=\phi(x)+\sum_n\eta_n(x)\cos(\omega_nt)$ into the equation of motion \eqref{eom}, where $\eta(x)$ is a small fluctuation around the static solution $\phi(x)$. This leads us to a Schr\"odinger-like equation,
\be\label{stabeq}
-\eta_n'' +U(x)\eta_n = \omega_n^2\eta_n,
\ee
where $U(x) = V_{\phi\phi} +\sigma(x)G_{\phi\phi}$ is the stability potential. If Eqs.~\eqref{VGW} and \eqref{fo} are valid, it reads
\be\label{U}
U(x) = W_{\phi\phi}^2 +W_\phi W_{\phi\phi\phi} +\sigma(x)W_{\phi\phi\phi}
\ee
and one can show that the zero mode exists, with the form $
\eta_0 = \phi'e^{-\int dx\frac{\sigma'}{\phi'}}$. Interestingly, the first-order formalism in Eqs.~\eqref{VGW} and \eqref{fo} allows for the factorization of the operator associated to Eq.~\eqref{stabeq} into the product of adjoint operators; see Ref.~\cite{impureza6}. The solution $\phi(x)$ is stable if negative eigenvalues are absent in \eqref{stabeq}. This occurs when the above zero mode is node-free.

\subsection{Impurity-free vacuumless systems}
Our interest is to investigate how to compactify the solutions that arise in vacuumless systems. Before dealing with this issue, let us review the impurity-free model presented in Refs.~\cite{vacuumless1,vacuumless2}. It is recovered with $\sigma(x)=0$ and
\be\label{wvac}
W(\phi) = \arctan(\sinh(\phi)).
\ee
In this situation, one gets the potential $V(\phi) = \frac12\sech^2(\phi)$, whose minima are located at $\phi\to\pm\infty$. The kink solution connecting them and its respective energy density can be obtained from Eqs.~\eqref{fo} and \eqref{rhow}, which lead to
\be\label{solvac}
\phi(x) = \arcsinh(x)\quad\text{and}\quad\rho(x) = \frac{1}{1+x^2}.
\ee
In this situation, the energy is $E=\pi$. The asymptotic behavior of these expressions can be written as
\be\label{solvacasy}
\phi(x) \approx \ln(2x)\quad\text{and}\quad\rho(x) \approx \frac{1}{x^2}.
\ee
The linear stability is described by the eigenvalue equation \eqref{stabeq} with the potential \eqref{U} under the condition $\sigma=0$, so we get
\be\label{uvac}
U(x) = \frac{2x^2-1}{(1+x^2)^2}.
\ee
It has a volcano profile, admitting only the zero mode $\eta_0(x)=(\sqrt{\pi}\sqrt{1+x^2})^{-1}$, whose absence of nodes confirms the stability of the impurity-free solution \eqref{solvac}.

Over the years, several papers dealing with the compactification of localized structures in impurity-free models appeared in the literature; see, for example, Refs.~\cite{comp1,comp2,comp23,comp24,kdvcomp1,kdvcomp2,kdvcomp3,comp222,comp3,comp4,comp45,comp5,k2c,comp6,comp7,comp8,comp89,comp899,comp8999,comp9,comp10,compschro}. However, the path to obtain compact solutions in vacuumless systems without impurities remains unknown. As a first tentative, one may try to construct a Lagrangian density \eqref{lmodel} with $\sigma=0$ supporting the expected behavior. For instance, one may start with the ansatz $\phi(x) = \tan(x)$ for $|x|<\pi/2$ and $\phi(x) = \text{sgn}(x)\,\infty$ for $|x|>\pi/2$, where $\text{sgn}(x)$ denotes the sign function. However, the energy of this tentative solution is infinite. This occurs due to the behavior $\phi(x)\propto1/(x\mp\pi/2)$ near the points $x=\pm\pi/2$ within the interval $|x|<\pi/2$. Since $\rho={\phi'}^2\propto1/(x\mp\pi/2)^4$ at the aforementioned points, we see that the energy is indeed infinite. For a second tentative, since the expected solution should present divergences at two points, $x=\pm \Tilde{x}$, we could suppose that $\phi(x)\propto1/(x\mp \Tilde{x})^a$, with $a>0$, near $x=\pm \Tilde{x}$ and try to adjust $a$ to obtain finite energy. This behavior appears in the energy density as $\rho(x)\propto a^2/(x\mp\Tilde{x})^{2+2a}$. To obtain an integrable divergence, we must impose $0<2+2a<1$, which is incompatible with $a>0$. Therefore, this tentative also fails. Actually, by using the Cauchy–Schwarz inequality, one can show that the energy obeys $E\geq\frac1{2b}\left(\int_{-b}^b\phi'\right)^2$, so we have
\be
    E\geq \frac1{2b}\left(\lim_{x\to b^-}\phi(x)-\lim_{x\to -b^+}\phi(x)  \right)^2,
\ee
where we have supposed that the compact vacuumless solution must obey $\phi(x)\to\infty$ ($\phi(x)\to-\infty)$ for $x\to b^{-}$ ($x\to -b^{+}$). The right hand side of the above inequality diverges for finite $b$, so the energy is infinite. Therefore, the impurity-free version of the model \eqref{lmodel} does not support such solutions. By using similar arguments, one can show that half-compact vacuumless solutions, which would obey  $\phi(x)\to\infty$ for $x\to b^{-}$ and $\phi(x)\to-\infty$ for $x\to-\infty$, do not appear in the aforementioned model.

\section{Presence of impurities}\label{secexamples}
Remarkably, we have found that the presence of impurities may lead to the compactification of the solutions in vacuumless systems. Considering the same auxiliary function in \eqref{wvac}, we can see that the inclusion of a non-null $\sigma(x)$ in the first-order equation \eqref{fo} gives us a hint about the compactification. Since the impurity is \emph{additive} in the aforementioned equation, we see that it may induce the solution to attain its boundary values nearer or farther to the origin. As our goal is to compactify the solution, we require the impurity to become larger and larger at specific points. 

\begin{figure}
\centering
\includegraphics[width=0.8\linewidth]{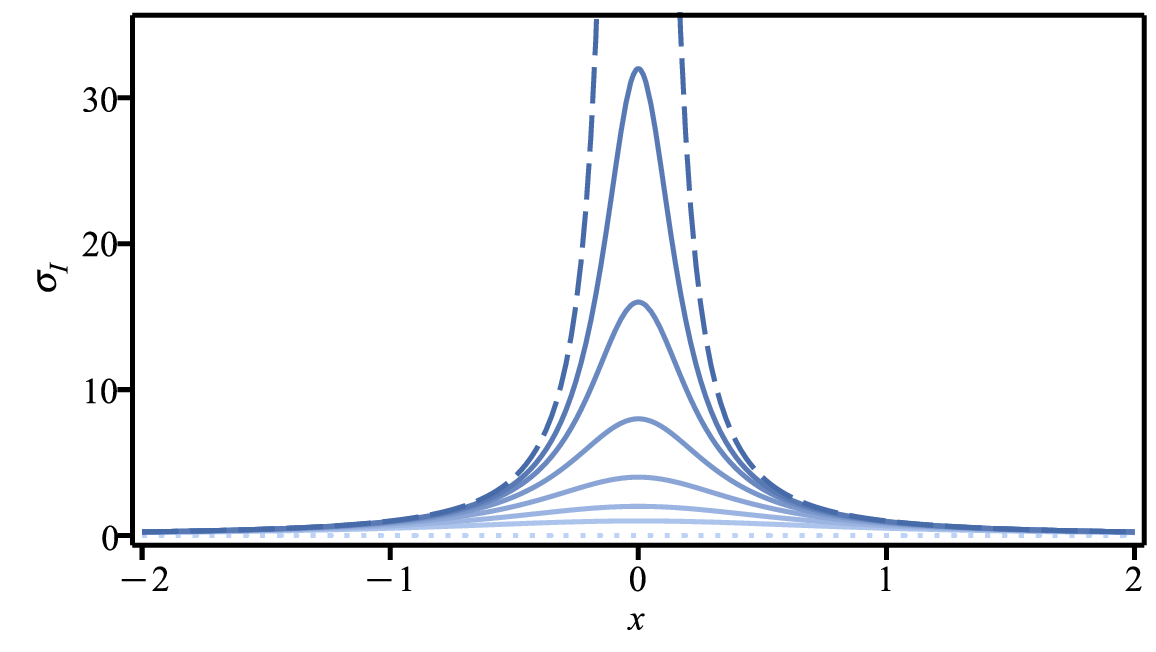}
\caption{The impurity \eqref{sigmaI} for $\alpha=0, 1, 2, 4, 8, 16, 32$ and $\alpha\to\infty$. The dotted line represents the case with $\alpha=0$, in which the impurity is absent, and the dashed one stands for the limit $\alpha\to\infty$, in which a divergence appears at the origin.}
\label{figsigmaI}
\end{figure}

To avoid breaking the monotonic character of the solutions, we consider non-negative impurities. First, we study an impurity with a single maximum located at the origin whose height is controlled by a non-negative parameter $\alpha$, in the form
\be\label{sigmaI}
\sigma_I(x) = \frac{\alpha}{1+\alpha x^2}.
\ee
Notice that the maximum obeys $\sigma_I(0)=\alpha$ and the asymptotic behavior of the above expression is $\sigma_I(x)\propto 1/x^2$, showing that it vanishes far away from the origin. The limit $\alpha\to\infty$ leads to
\be\label{sigmaIinf}
\sigma^\infty_I(x) = \frac{1}{x^2},
\ee
which engenders a divergence at the origin. In Fig.~\ref{figsigmaI}, we display the behavior of the impurity \eqref{sigmaI} for several values of $\alpha$, including the limit $\alpha\to\infty$ given above. One notes that $\sigma_I(x)$ represents a symmetric Lorentzian function, which is known to appear in spectroscopy, where it may engender fundamental signatures that are used to quantify structural and dynamical properties of nuclei, atoms, and molecules.

\begin{figure}
\centering
\includegraphics[width=0.333\linewidth]{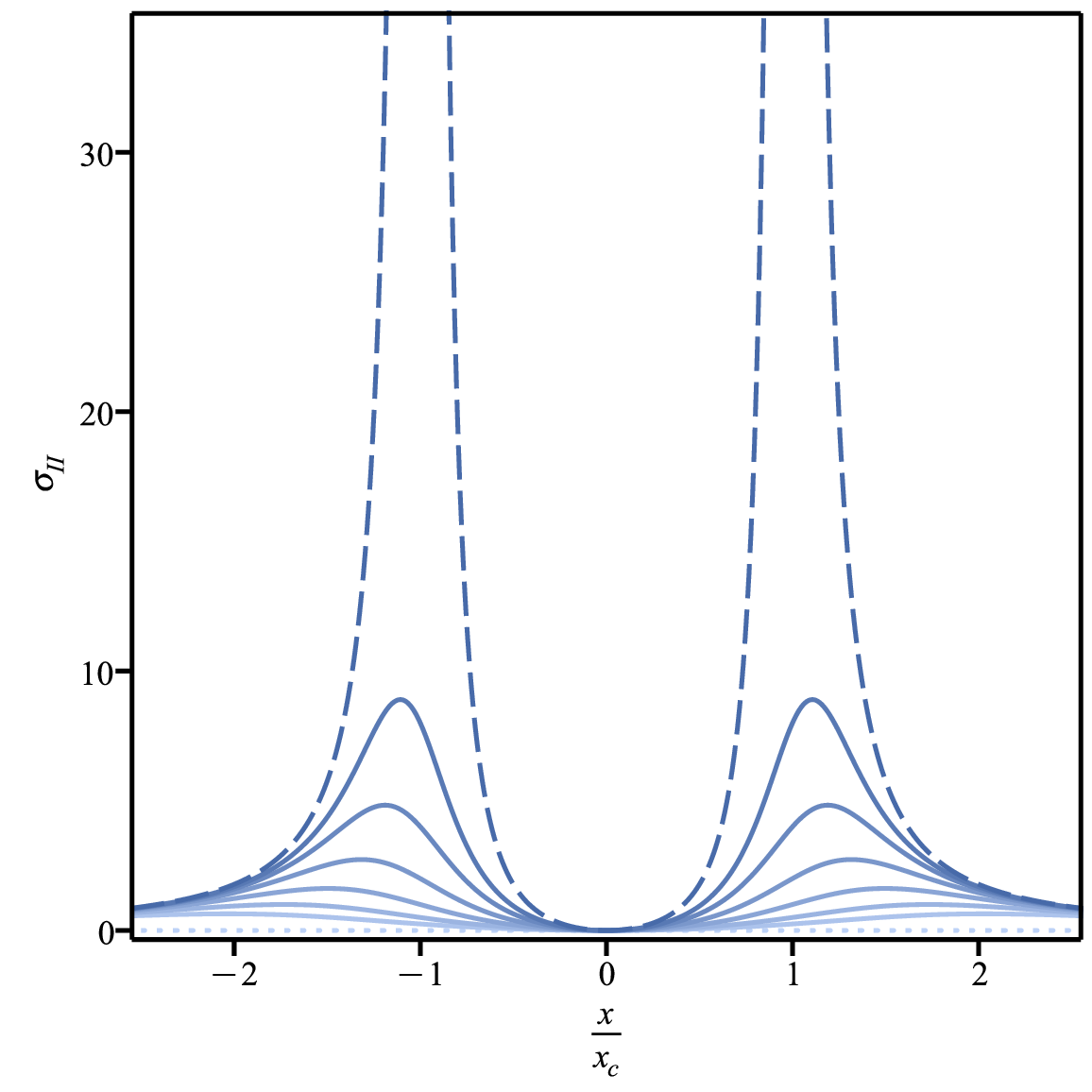}\includegraphics[width=0.333\linewidth]{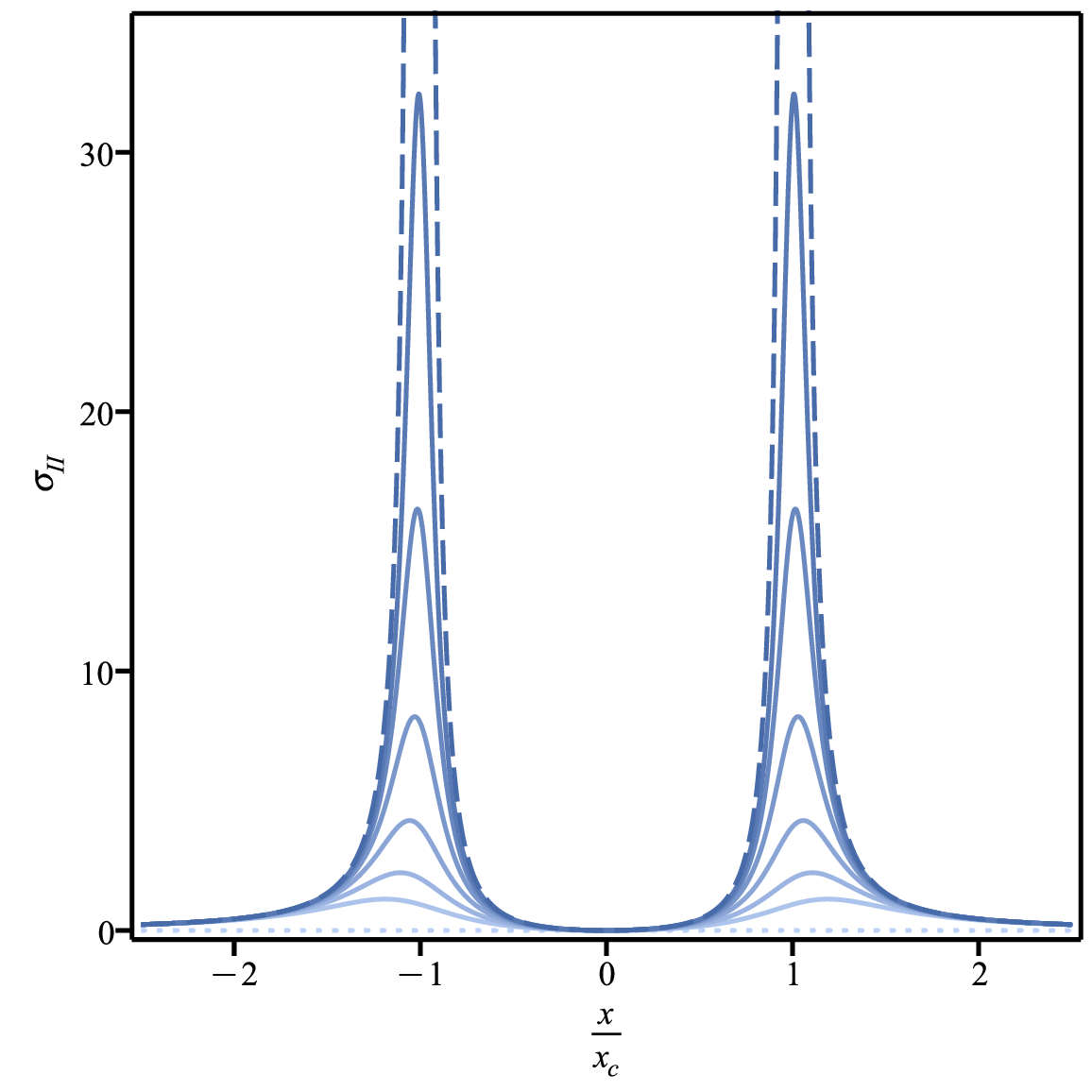}\includegraphics[width=0.333\linewidth]{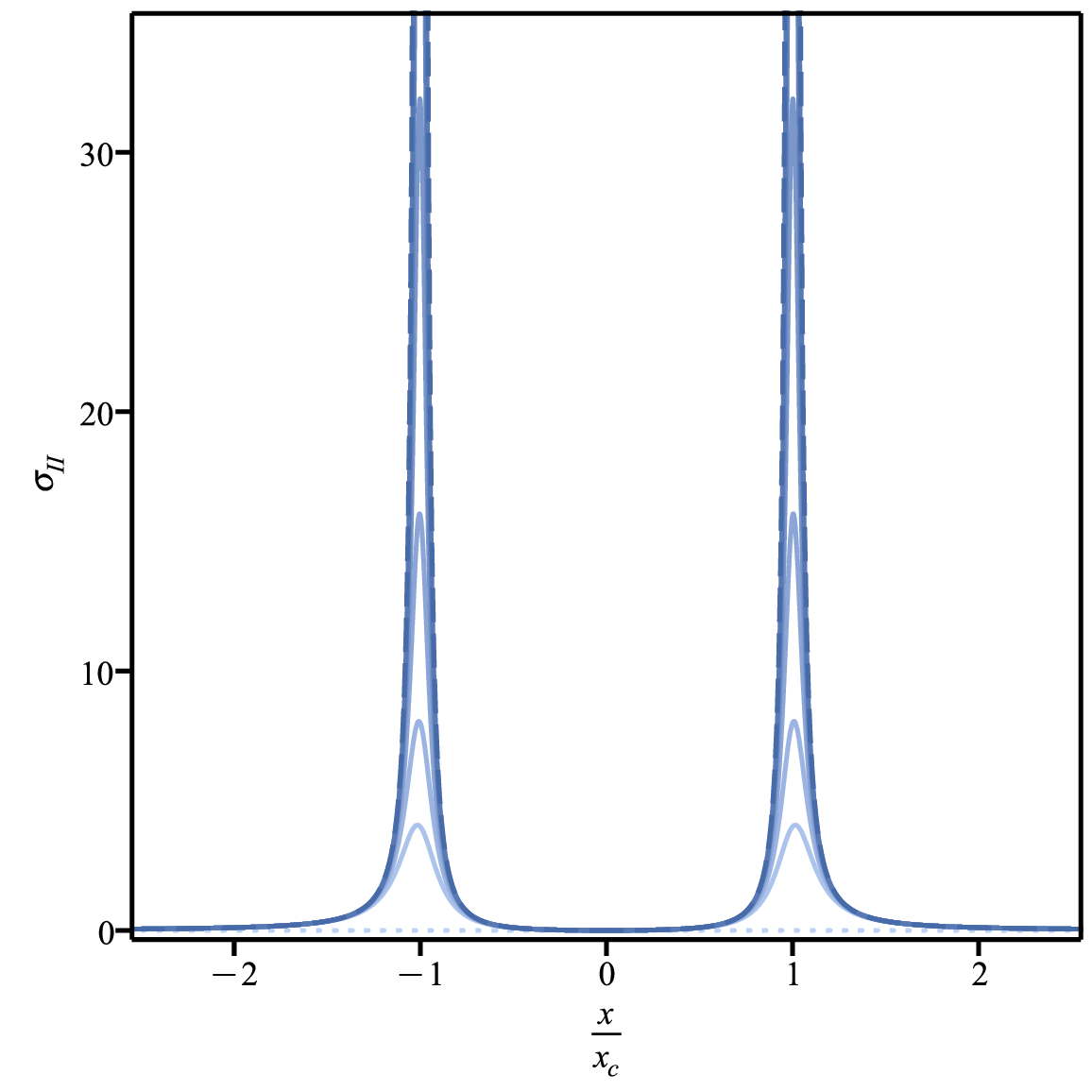}
\caption{The impurity \eqref{sigmanum1} with respect to $x/x_c$ for $\alpha=0, 1, 2, 4, 8, 16, 32$ and $\alpha\to\infty$. The left, middle and right panels stand for $x_c=1/2,1$ and $2$. The dotted lines represent the case with $\alpha=0$ and the dashed ones stand for the limit $\alpha\to\infty$.}
\label{figsigma1}
\end{figure}

We also explore another possibility, in which the maxima of the impurity occur at two equidistant points around the origin. In this situation, the parameter $\alpha$, which we suppose to be non negative, controls the location and the height of the maxima. We then write
\be\label{sigmanum1}
\sigma_{II}(x) = \frac{\alpha\,x^2}{1+\alpha\left(x_c^2-x^2\right)^2},
\ee
where $x_c$ is a real parameter. Notice that $\alpha=0$ recovers the impurity-free scenario, in which $\sigma=0$. For $\alpha$ positive, this function engenders two maxima located at $x_m=x_c(1+(\alpha x_c^4)^{-1})^{1/4}$, such that $\sigma(x_m) = \sqrt{\alpha}(\sqrt{\alpha}x_c^2+\sqrt{1+\alpha x_c^4})/2$. Asymptotically, we have $\sigma(x)\propto 1/x^{2}$, so the above expression vanishes for $x\to\infty$, showing that this impurity is spatially localized. The limit with infinite $\alpha$ leads to
\be\label{sigmacomp1}
\sigma^\infty_{II}(x) = \frac{x^2}{\left(x_c^2-x^2\right)^2}.
\ee
In this limit, we have $x_m=x_c$ and the maxima become divergent points. Notice that the above expression becomes the very same in Eq.~\eqref{sigmaIinf} for $x_c=0$. In Fig.~\ref{figsigma1}, we display the impurity \eqref{sigmanum1} for some values of $x_c$ and $\alpha$, including the limit $\alpha\to\infty$ given above.

Next, we investigate the behavior of the field configurations with the impurities \eqref{sigmaI} and \eqref{sigmanum1}.

\subsection{Presence of $\sigma_I(x)$}\label{secsigmaI}
We couple \eqref{sigmaI} to the scalar field model described by \eqref{wvac} via the Lagrangian density \eqref{lmodel}. Using the first-order equation \eqref{fo}, we get
\be\label{foI}
\phi'= \sech(\phi) + \frac{\alpha}{1+\alpha x^2}.
\ee
This expression allows us to see that the slope of the solution never vanishes or becomes negative, which is the same behavior as the impurity-free solution \eqref{solvac}. Using the above equation, the energy density in Eq.~\eqref{rhow} can be written in the form
\be\label{rhoI}
\rho(x) = \left(\sech(\phi(x)) + \frac{\alpha}{1+\alpha x^2}\right)\sech(\phi(x))
\ee
and the stability potential \eqref{U} as
\be\label{UI}
\begin{aligned}
U(x) &= \left(2-3\,\sech^2(\phi(x))\right)\sech^2(\phi(x))\\
    &+\frac{\alpha\left(1-2\,\sech^2(\phi(x))\right)\sech(\phi(x))}{1+\alpha x^2},
\end{aligned}
\ee
where $\phi(x)$ is the solution of \eqref{foI}.

\begin{figure}
\centering
\includegraphics[width=0.333\linewidth]{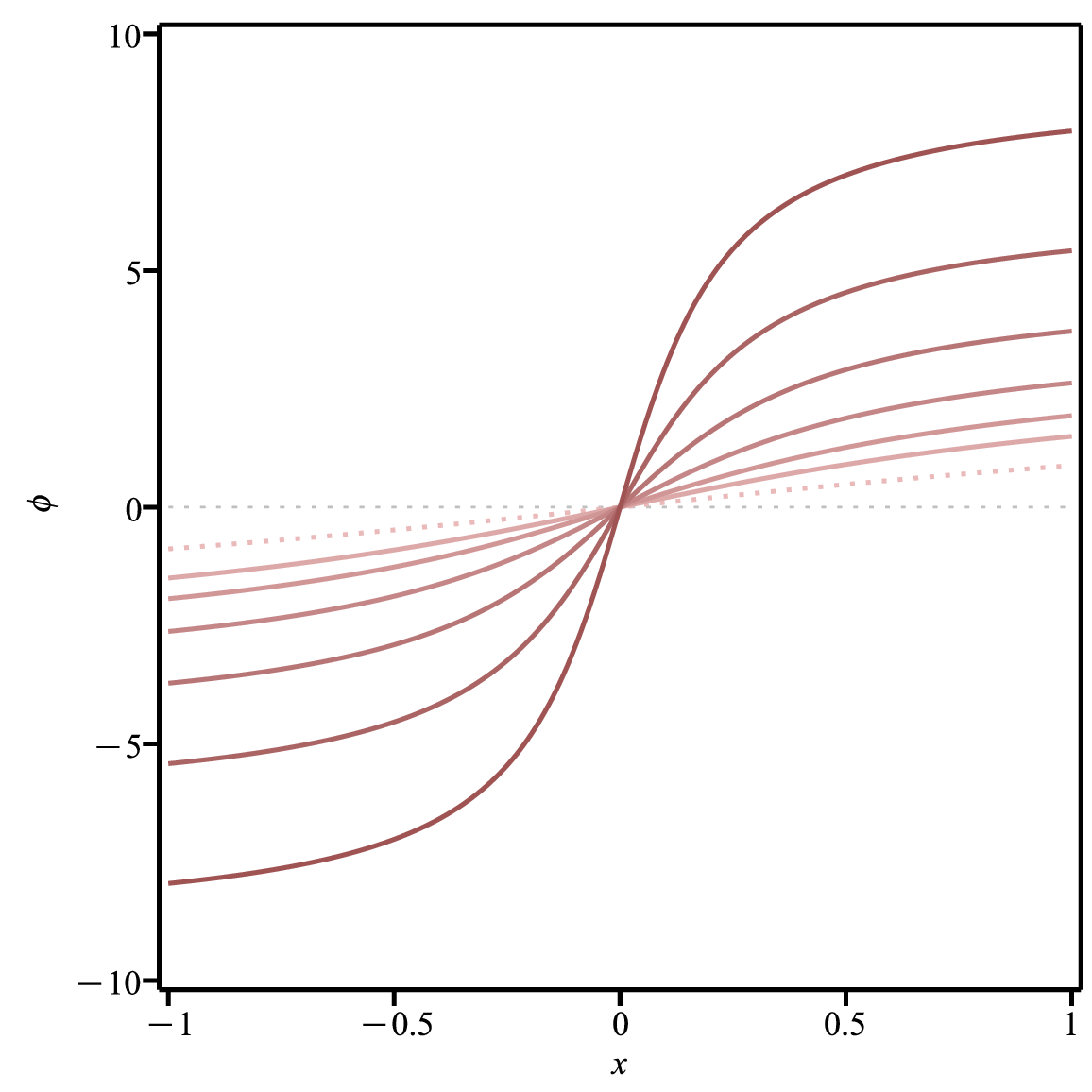}\includegraphics[width=0.333\linewidth]{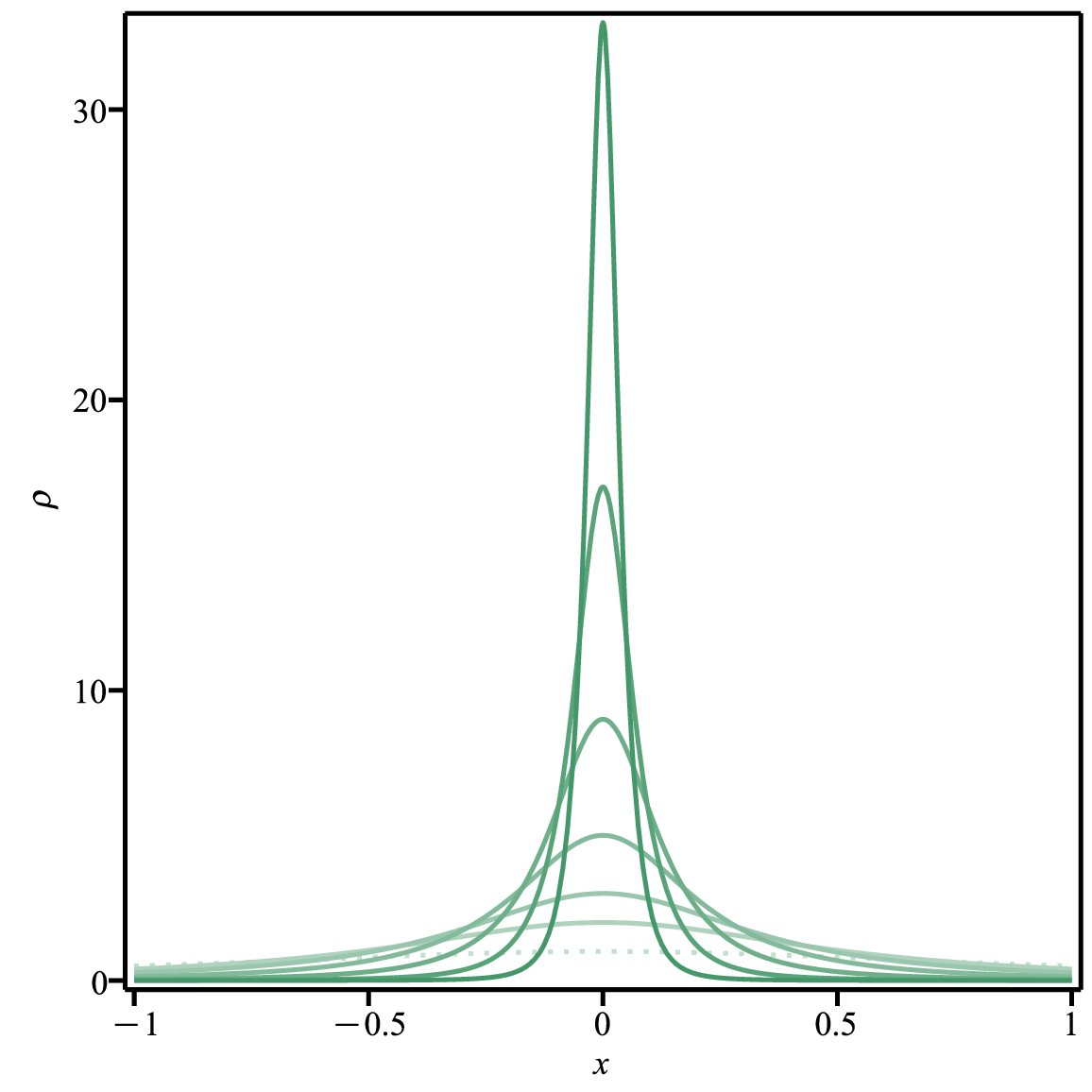}\includegraphics[width=0.333\linewidth]{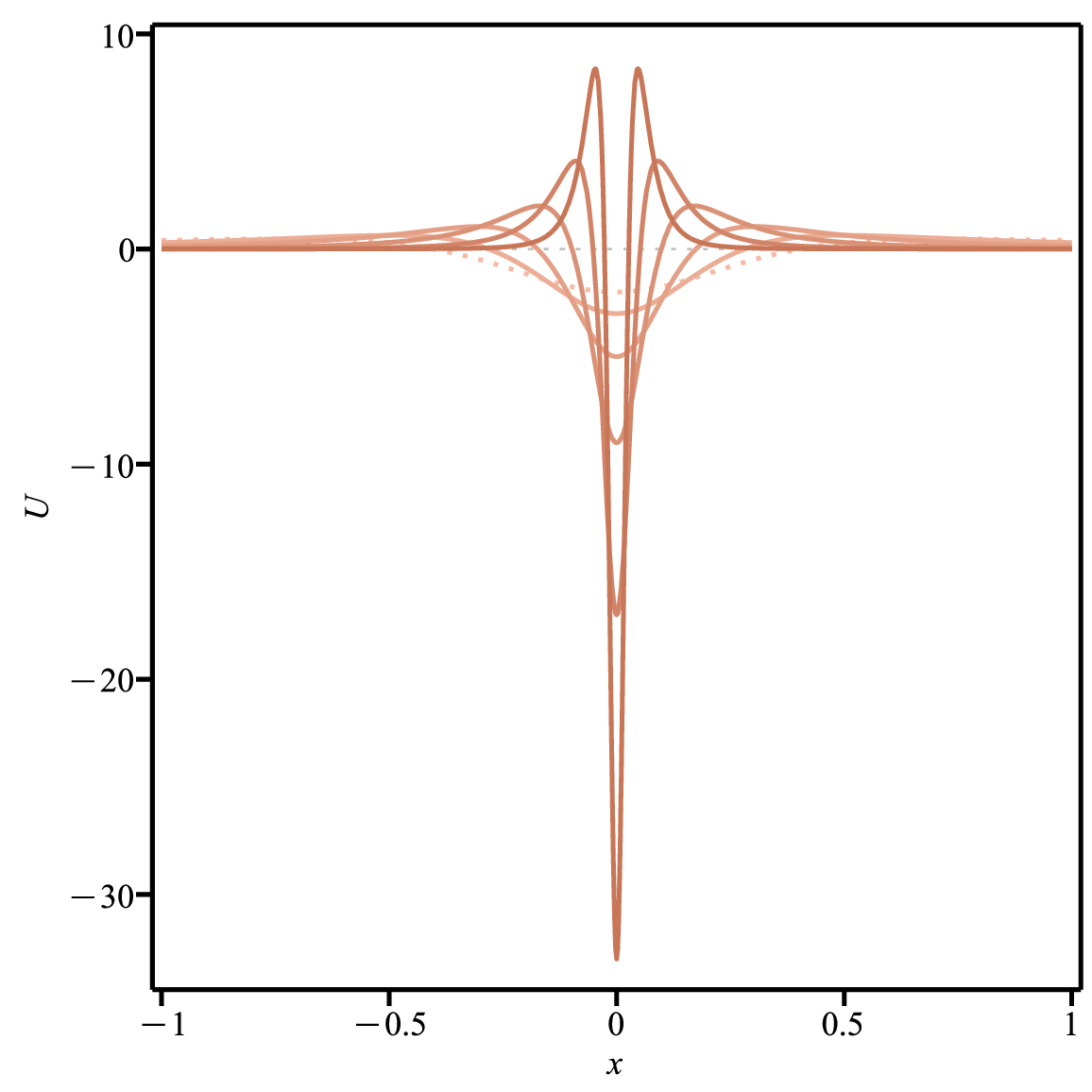}
\includegraphics[width=0.333\linewidth]{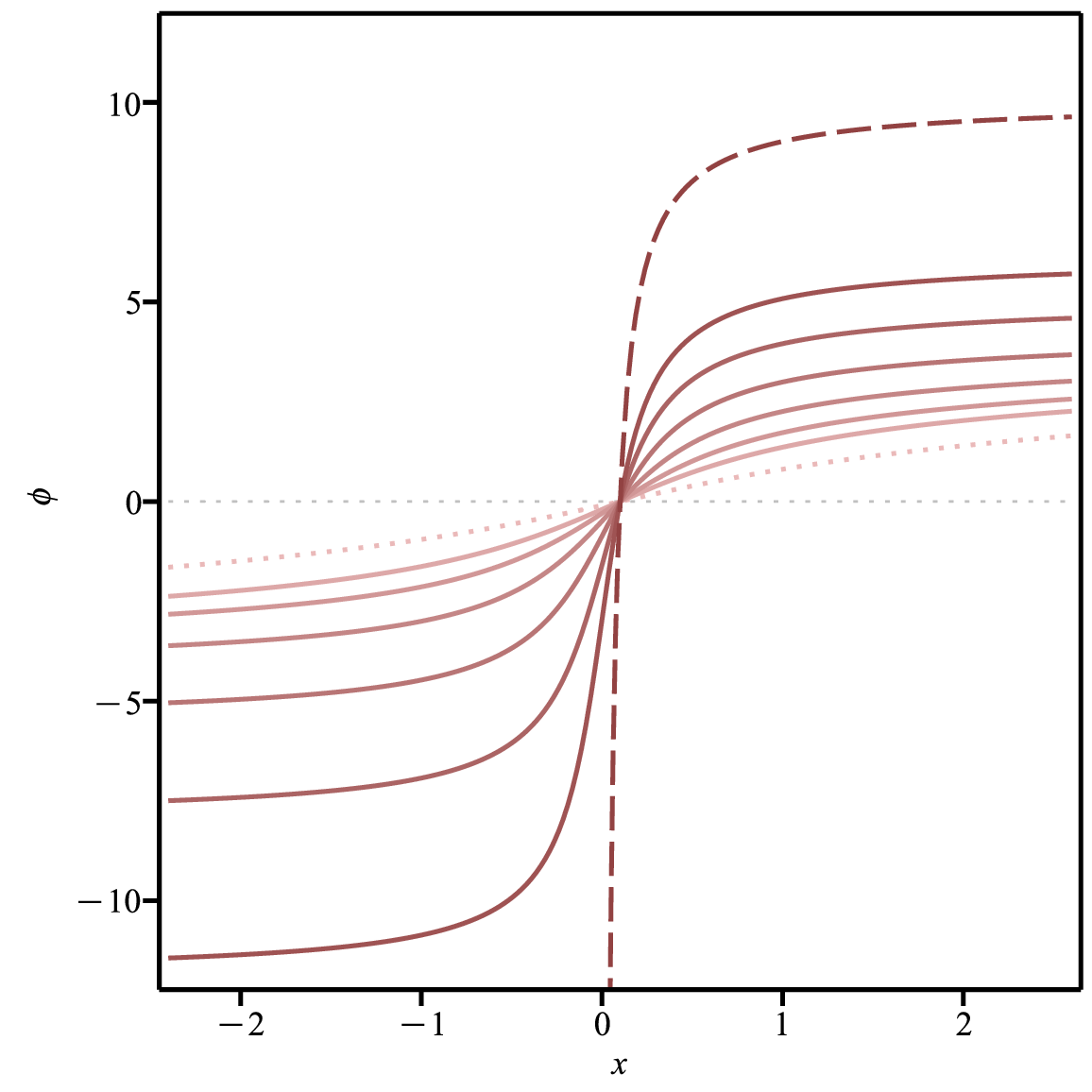}\includegraphics[width=0.333\linewidth]{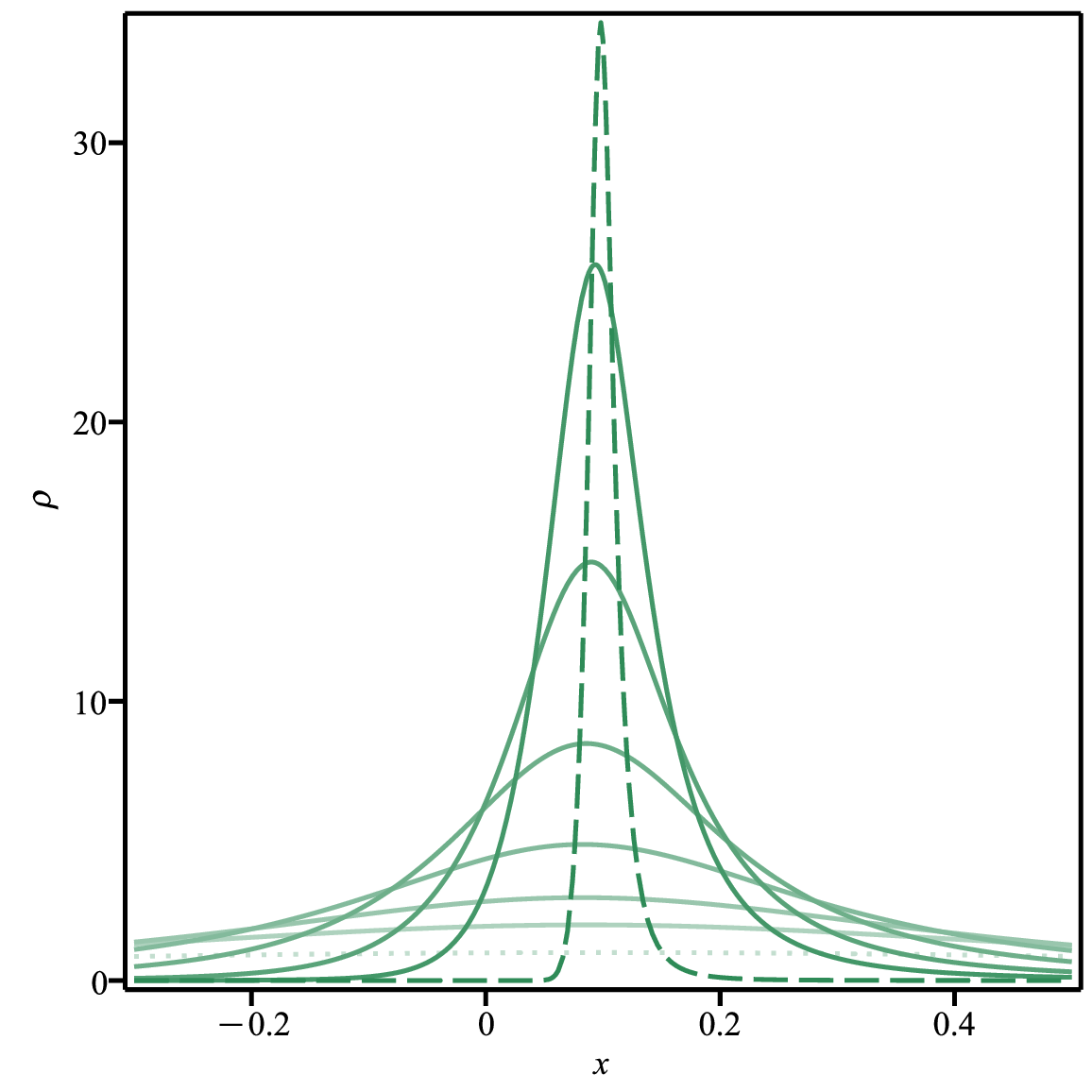}\includegraphics[width=0.333\linewidth]{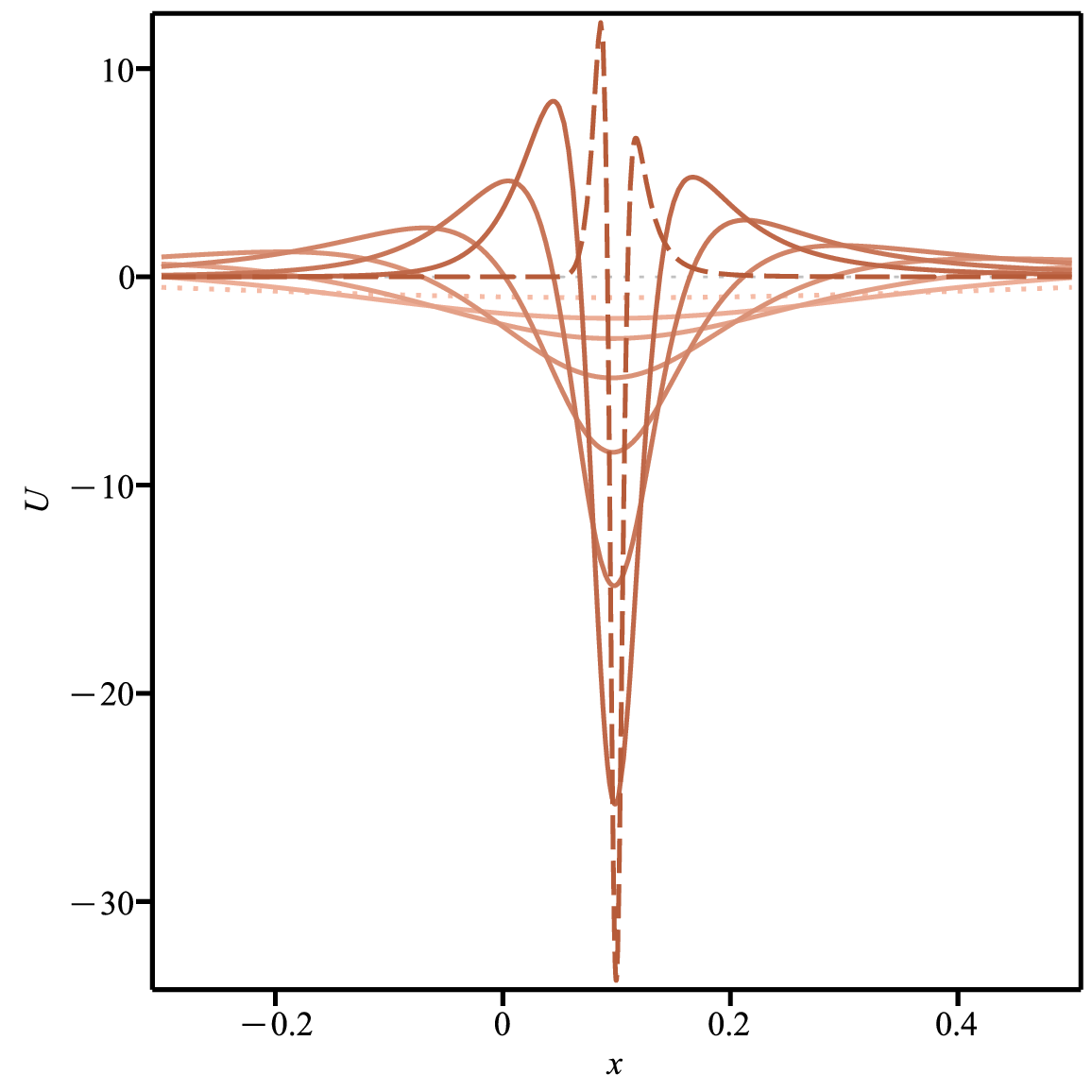}
\includegraphics[width=0.333\linewidth]{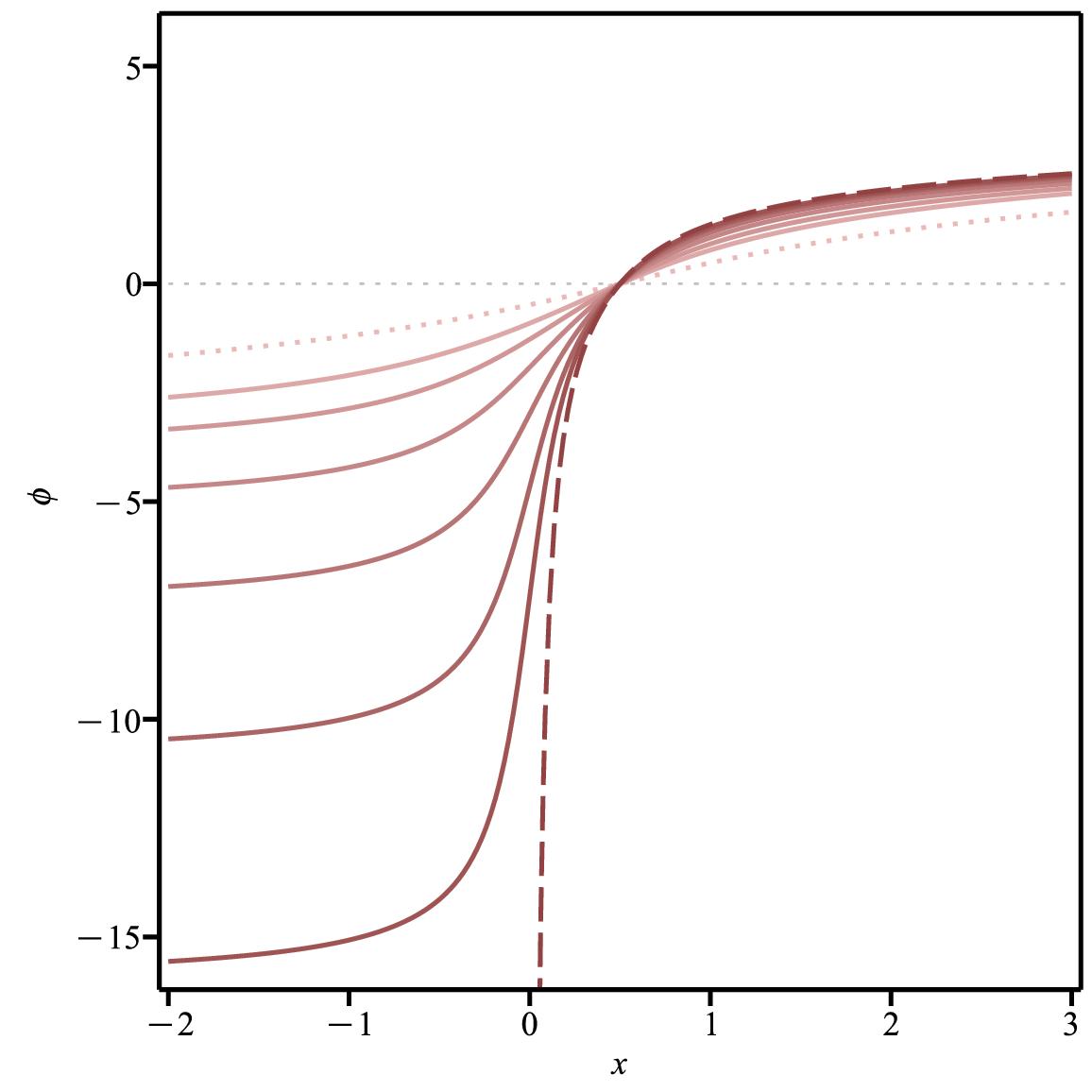}\includegraphics[width=0.333\linewidth]{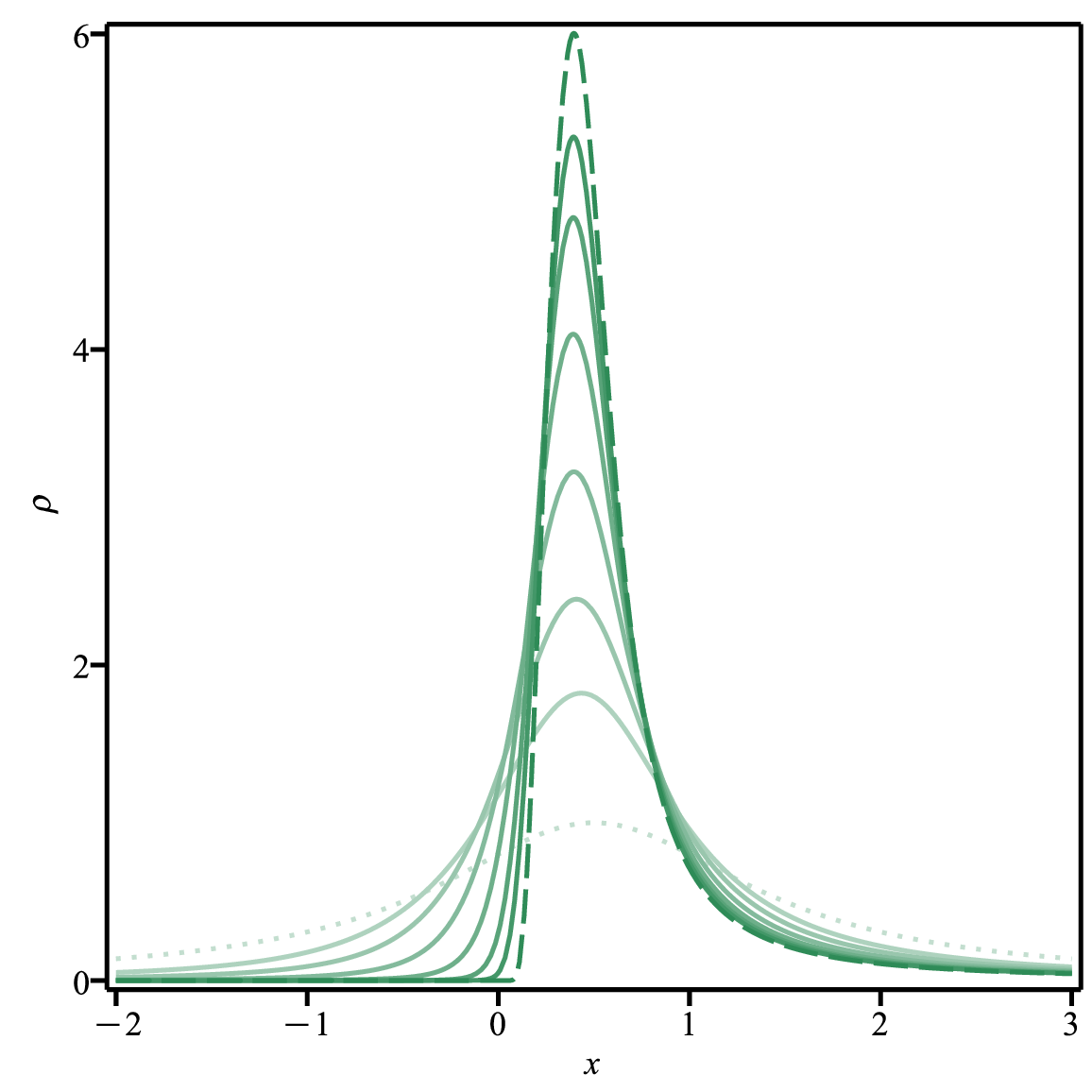}\includegraphics[width=0.333\linewidth]{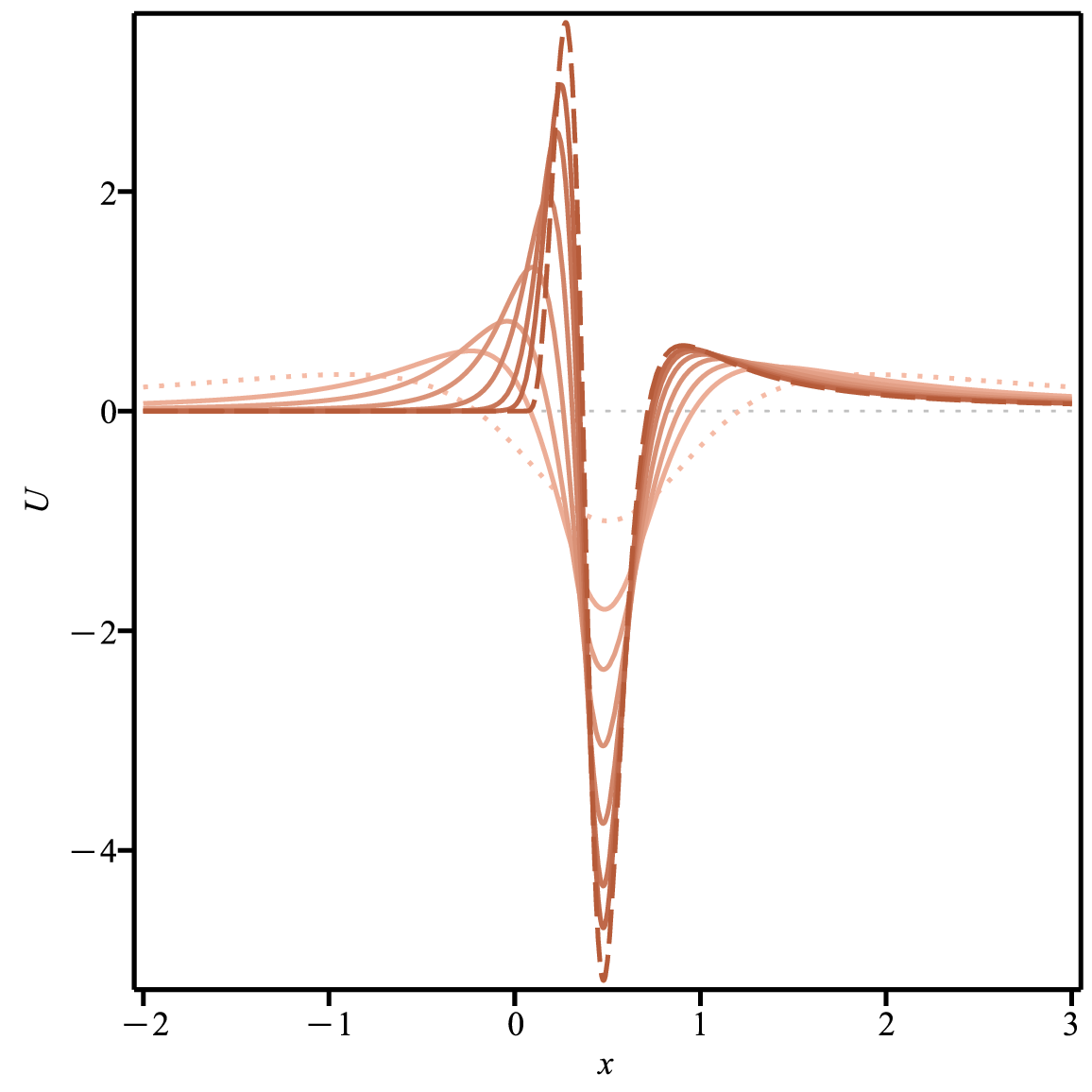}
\caption{The solution $\phi(x)$ of the first-order equation \eqref{foI} (left) calculated with the condition $\phi(x_0)=0$, its energy density \eqref{rhoI} (middle) and the stability potential \eqref{UI} (right) for the same line styles and values of the parameter $\alpha$ in Fig.~\ref{figsigmaI}, with $x_0=0$ (top panels), $0.1$ (middle panels) and $0.5$ (bottom panels). To provide a better visualization, we plotted $\rho(x)/3$ and $U(x)/3$ for the case with $x_0=0.1$ and $\alpha\to\infty$.}
\label{fignumI}
\end{figure}

For $\alpha=0$, we recover the results in \eqref{solvac} and \eqref{uvac}. However, the inhomogeneity introduced by the impurity in the first-order equation hinders us to get the analytical solution for $\alpha$ positive, so numerical procedures are required. Before using them, let us investigate the behavior of the solution at specific points in the simplest situation, which occurs for the condition $\phi(0)=0$, preserving the odd symmetry in $\phi(x)$. Near the origin, we get $\phi(x)\approx (1+\alpha)x$. Therefore, the slope of the solution increases as $\alpha$ also increases. The asymptotic behavior is the same in Eq.~\eqref{solvacasy}. In the top panels of Fig.~\ref{fignumI}, we display the symmetric solution of \eqref{foI}, the energy density \eqref{rhoI} and the stability potential \eqref{UI} for several values of $\alpha$. As $\alpha$ increases, we see that, at the origin, the slope of $\phi(x)$ becomes larger, the energy density gets taller and the volcano stability potential attains a deeper valley. The limit $\alpha\to\infty$ brings an interesting feature to light: the solution tends to the singular kink $\phi(x)=\text{sgn}(x)\,\infty$, with energy density $\rho(x) = \pi\delta(x)$, where $\delta(x)$ represents the Dirac delta function. This solution has the same form of the tachyon kink found in Refs.~\cite{asen1,asen2,trilogia1}.

The presence of the impurity \eqref{sigmaI} also motivates us to investigate how the solution of \eqref{foI} behaves under the condition $\phi(x_0)=0$. Near the point $x=x_0$, the behavior is $\phi(x) \approx (1+\sigma_I(x_0))(x-x_0)$ so the slope of the solution increases with $\alpha$ but remains finite for $\alpha\to\infty$. On the other hand, the asymptotic behavior is still the same in Eq.~\eqref{solvacasy}. Notice that, differently from the symmetric case, the presence of a non-null $x_0$ makes the peak of the impurity acts on $x=0$, which is does not coincide with $x_0$, increasing the slope at the origin as $\alpha$ gets larger. In this situation, the tail of the solution tends to compactify at the origin. The limit $\alpha\to\infty$ leads to a half-compact profile. In the middle- and bottom-row panels of Fig.~\ref{fignumI}, we show the solution, energy density and stability potential for $x_0=0.1$ and $0.5$, respectively. In the limit $\alpha\to\infty$, the left tail of the solution is compactified so, for $x\leq x_0$, $\phi(x)\to-\infty$, $\rho(x)=0$ and $U(x)=0$. This reveals a novel structure, engendering half-compact profile, in impurity-doped vacuumless systems. Notice that, for small values of $x_0$, the energy density's maximum gets large values and and the stability potential's valley becomes deep and narrow. 

\subsection{Presence of $\sigma_{II}(x)$}
The impurity \eqref{sigmanum1} makes the first-order equation \eqref{fo} associated to the vacuumless system \eqref{wvac} become
\be\label{fonum}
\phi' = \sech(\phi) + \frac{\alpha\,x^2}{1+\alpha\left(x_c^2-x^2\right)^2}.
\ee
Similarly to what occurs with \eqref{foI}, the above expression has a positive right-hand side, leading to monotonic solutions. It allows us to calculate the energy density \eqref{rhow} in a simpler form which, for the auxiliary function \eqref{wvac} and the impurity \eqref{sigmanum1}, becomes
\be\label{rhonum1}
\rho(x) = \left(\sech(\phi(x)) + \frac{\alpha\,x^2}{1+\alpha\left(x_c^2-x^2\right)^2}\right)\sech(\phi(x)),
\ee
where $\phi(x)$ denotes the solution of the first-order equation \eqref{fonum}. Also, the stability potential \eqref{U} reads
\be\label{Unum1}
\begin{aligned}
U(x) &= \left(2-3\,\sech^2(\phi(x))\right)\sech^2(\phi(x))\\
    &+\frac{\alpha\,x^2\left(1-2\,\sech^2(\phi(x))\right)\sech(\phi(x))}{1+\alpha\left(1-x^2\right)^2}.
\end{aligned}
\ee
To calculate $\rho(x)$ and $U(x)$, of course, one must solve the first-order equation \eqref{fonum}. For $\alpha=0$, we recover the results in \eqref{solvac} and \eqref{uvac}. However, for positive $\alpha$, we must use numerical methods.

Solutions obeying $\phi(0)=0$ preserves the odd symmetry. In this situation, near the origin, $x\approx0$, Eq.~\eqref{fonum} leads us to $\phi(x)\approx x$, which does not depend on $\alpha$ so, contrary to the impurity \eqref{sigmaI}, the presence of \eqref{sigmanum1} does not introduce changes around this point. Asymptotically, one can show that it has the very same form in \eqref{solvacasy}, which also does not depend on $\alpha$. The impact of $\alpha$ on the solution, therefore, is outside the origin, at the points $x=\pm x_m$ as in the expression below Eq.~\eqref{sigmanum1}. For finite $\alpha$, the solution is extended, spanning from $-\infty$ to $\infty$ as $x$ ranges from $-\infty$ to $\infty$, as in the impurity-free case. As $\alpha$ increases, the derivative of the solution gets larger at the maxima of the impurity \eqref{sigmanum1}. In the limit $\alpha\to\infty$, the slope of the solution tends to infinity at $x=\pm x_c$, giving rise to \emph{compact} structures. Indeed, for points near $x=\pm x_c$ inside the interval $|x|<x_c$, the solution behaves as
\be
\phi(x)\propto \frac1{x_c\mp x}.
\ee
To preserve continuity, we have that $\phi(x)\to\text{sgn}(x)\infty$ outside the limited interval $|x|< x_c$. 

\begin{figure}
\centering
\includegraphics[width=0.333\linewidth]{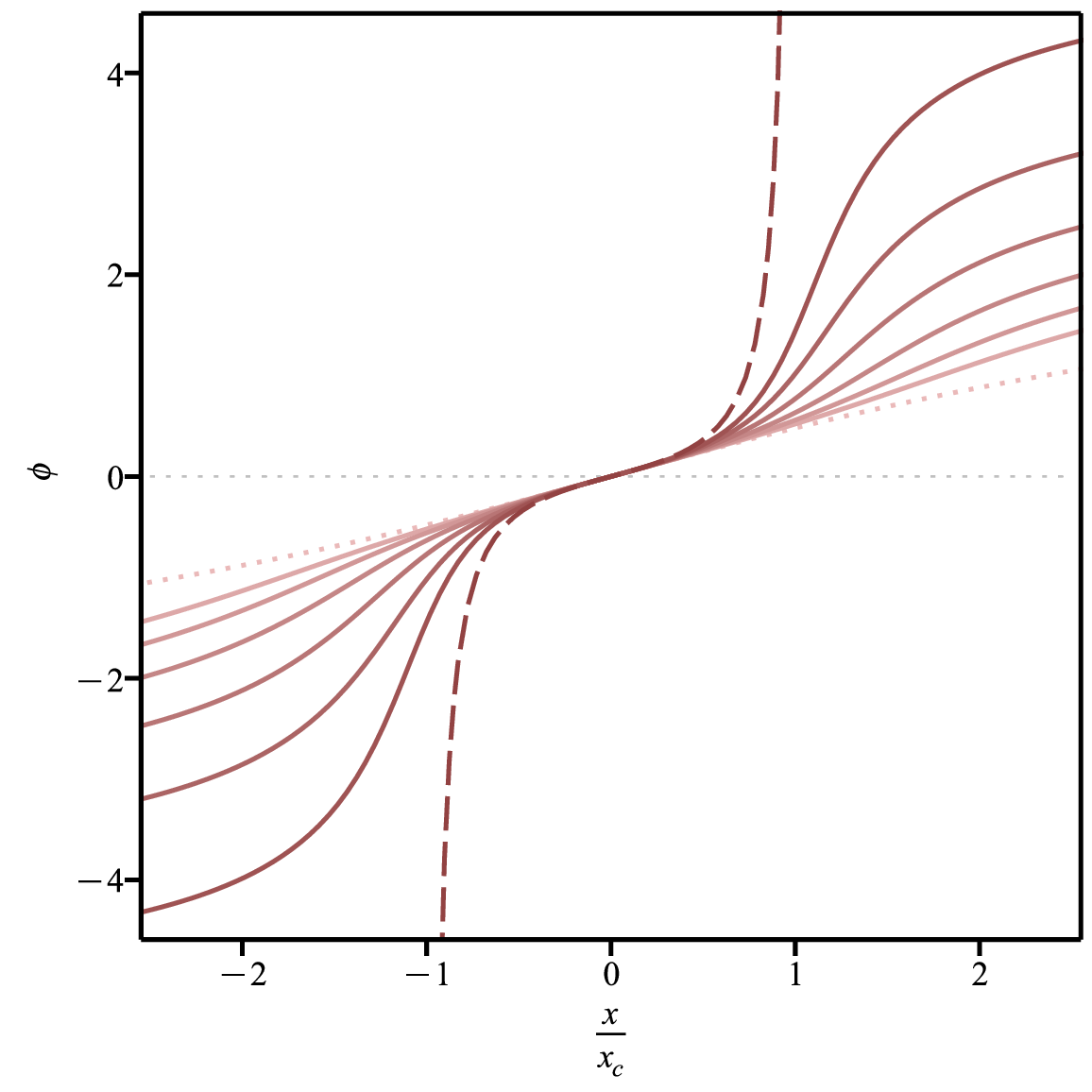}\includegraphics[width=0.333\linewidth]{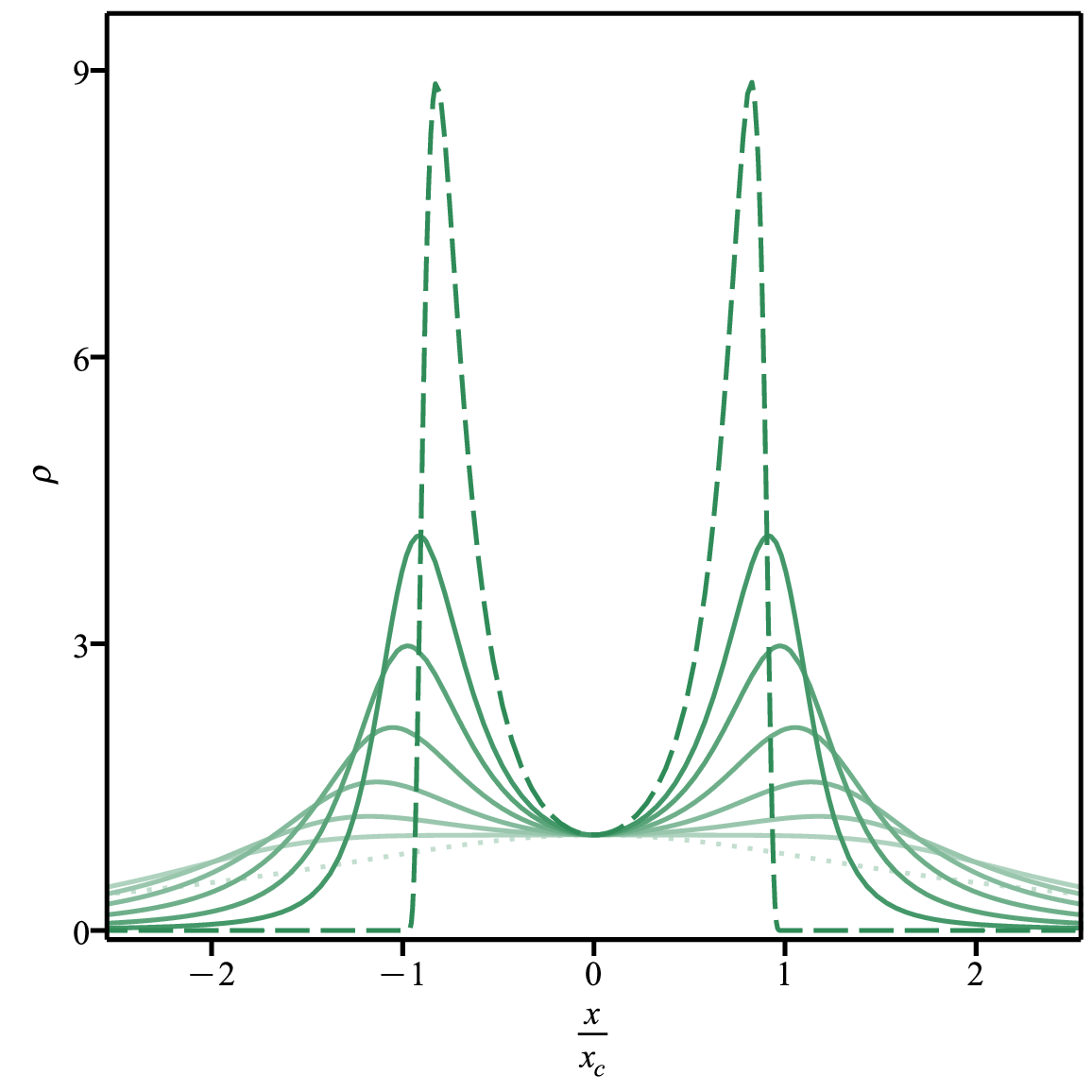}\includegraphics[width=0.333\linewidth]{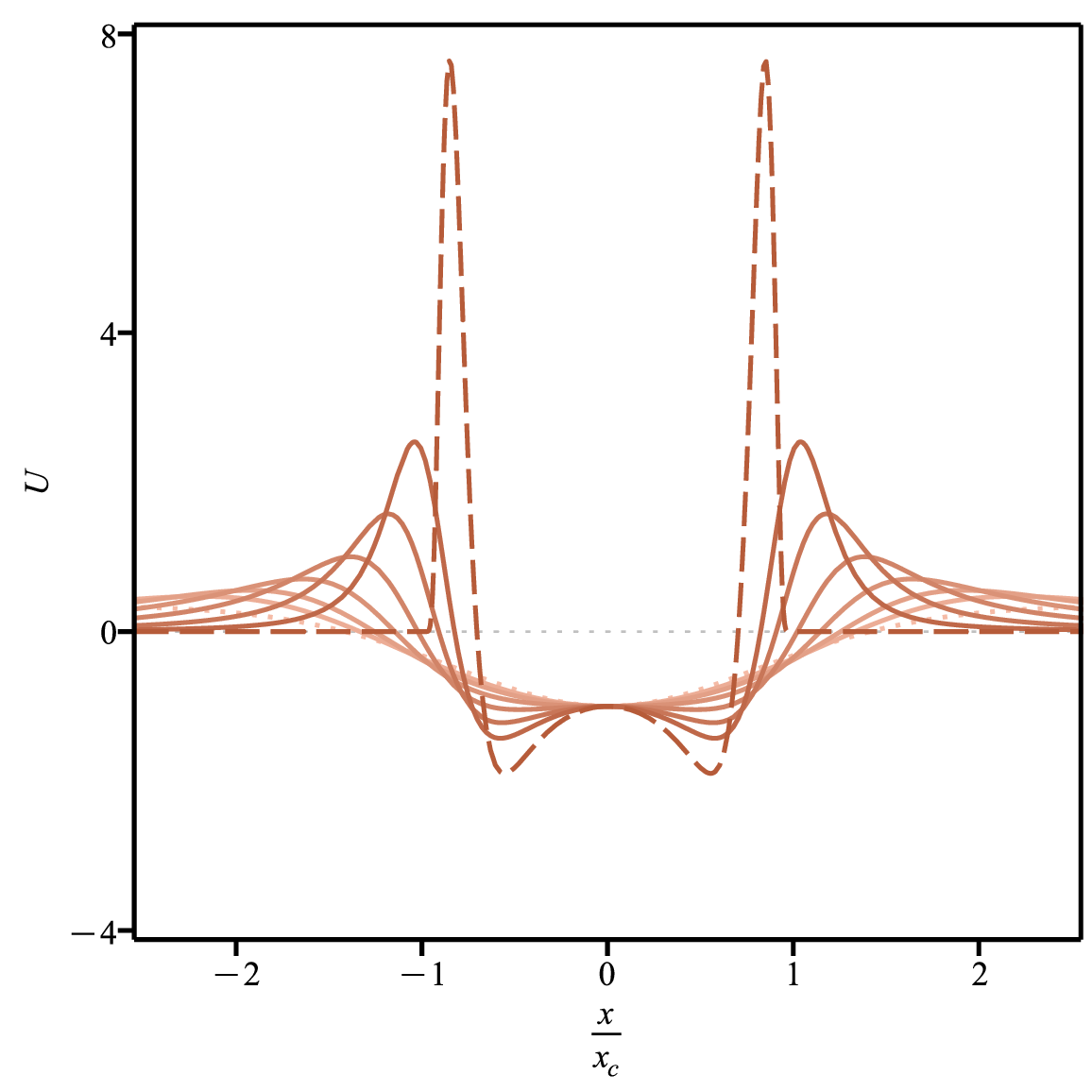}
\includegraphics[width=0.333\linewidth]{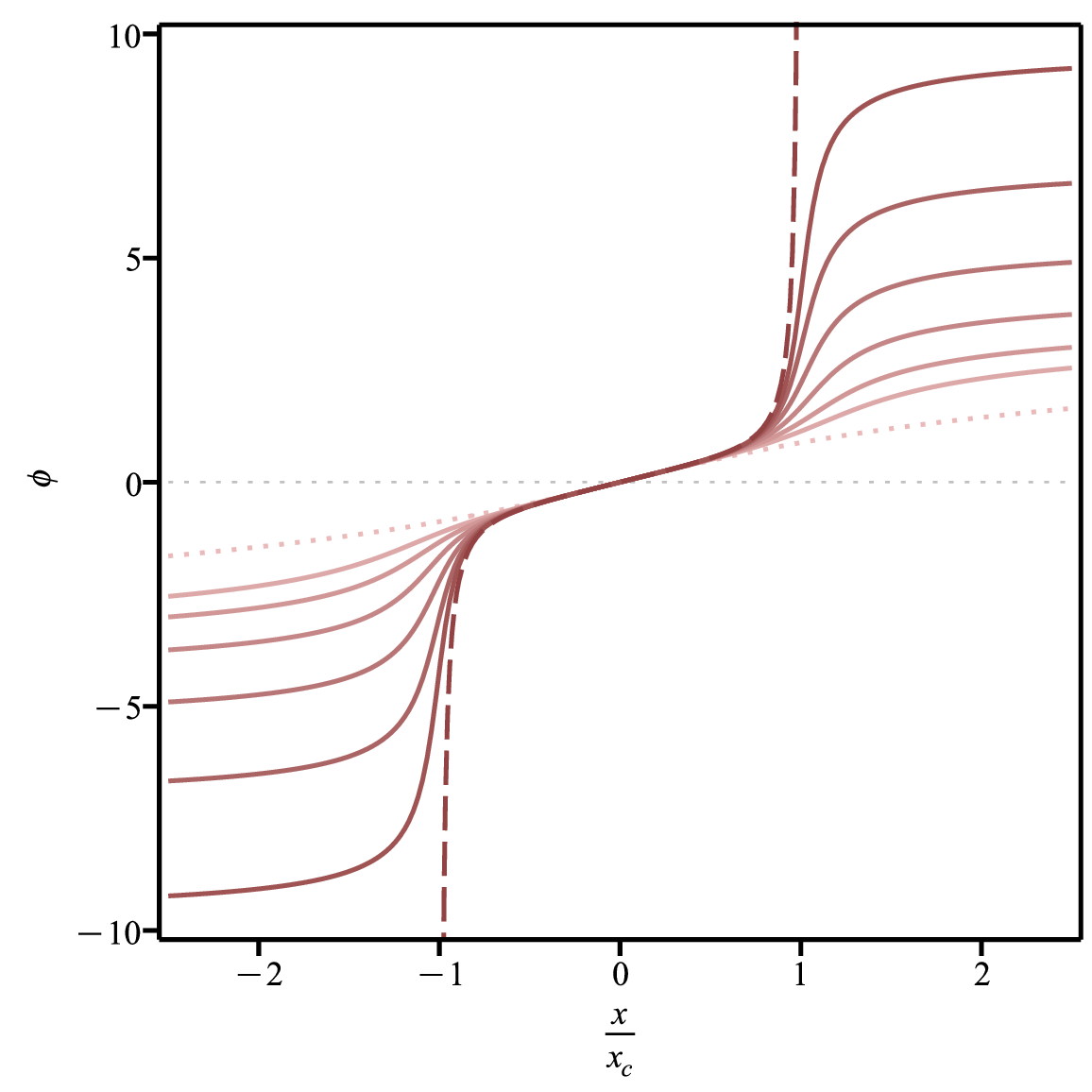}\includegraphics[width=0.333\linewidth]{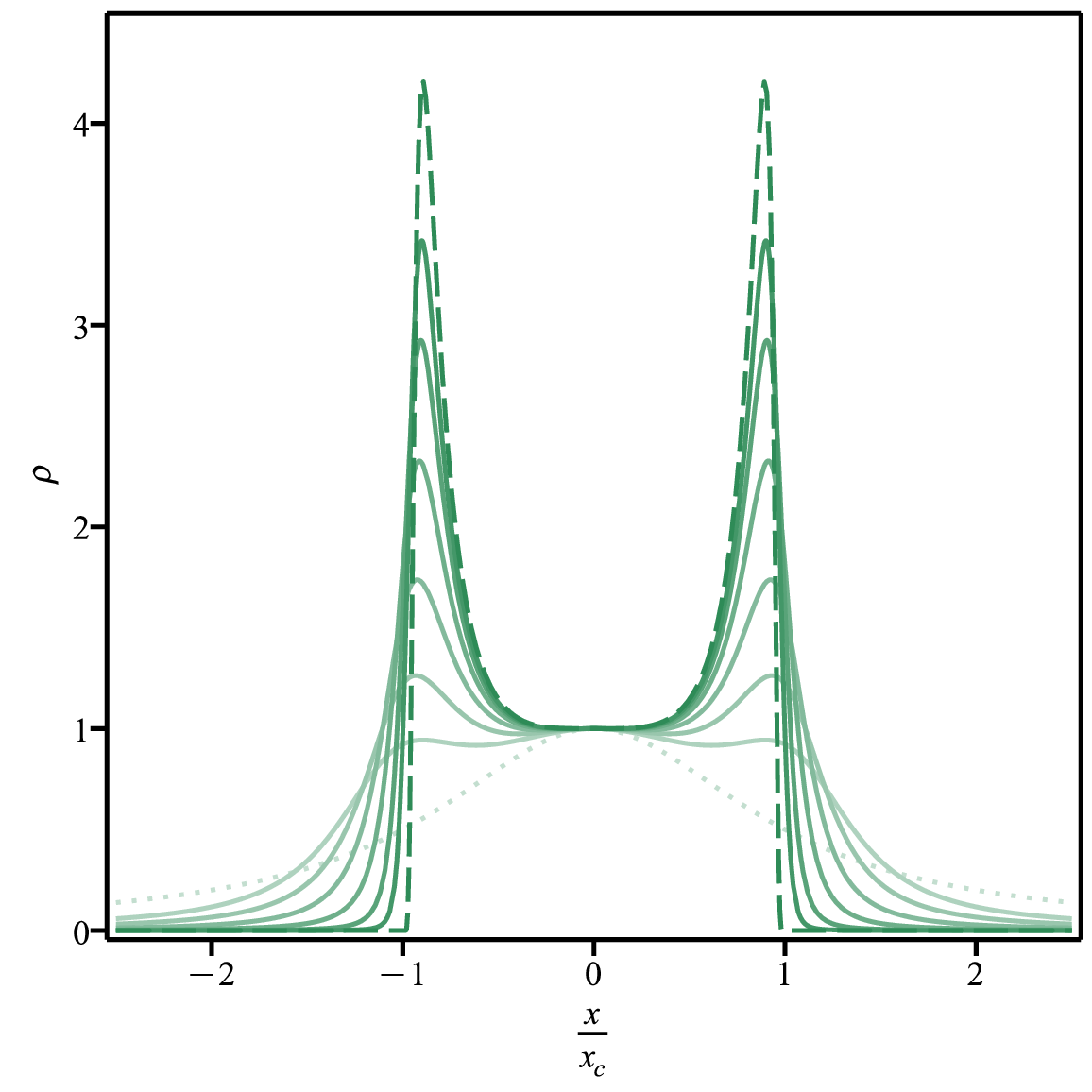}\includegraphics[width=0.333\linewidth]{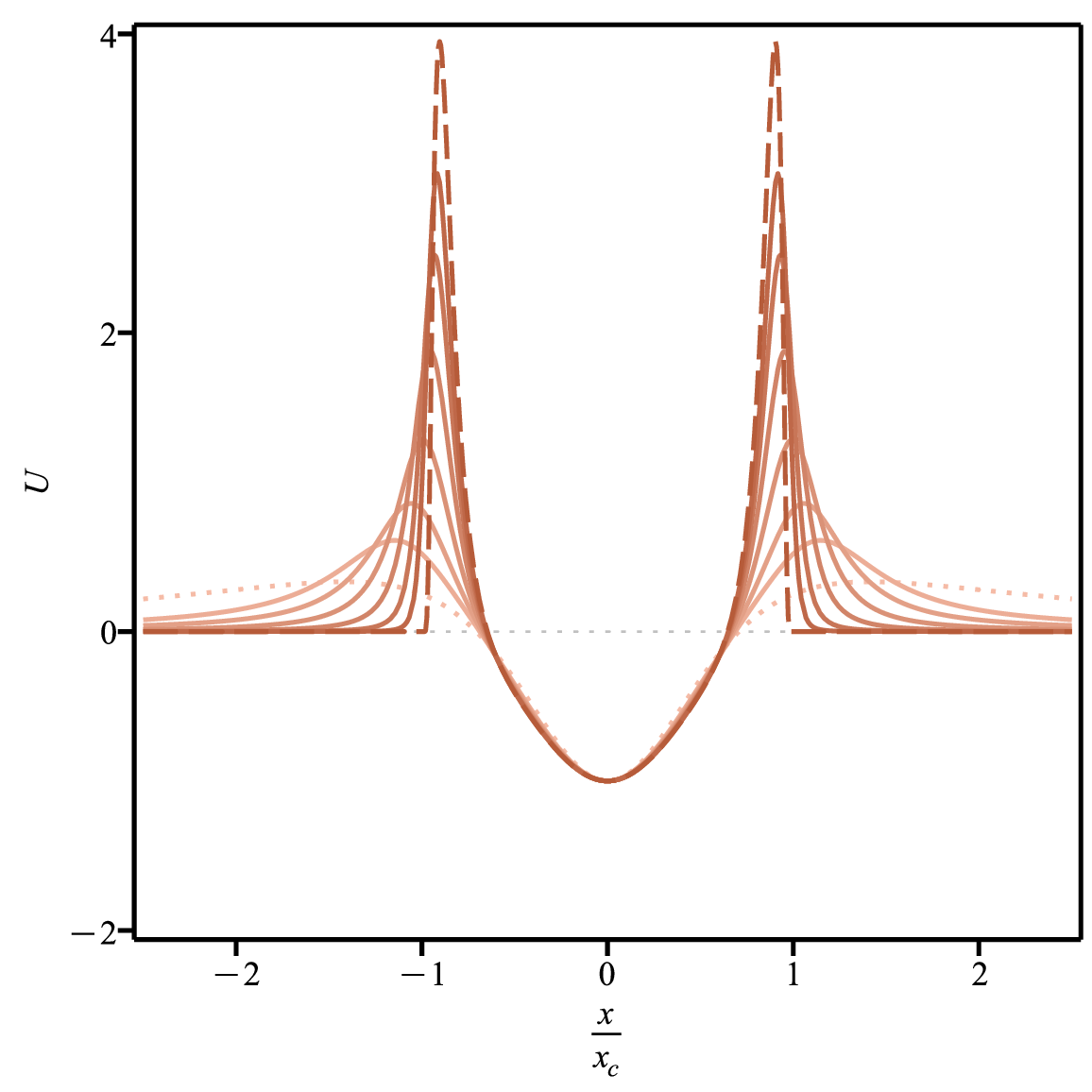}
\includegraphics[width=0.333\linewidth]{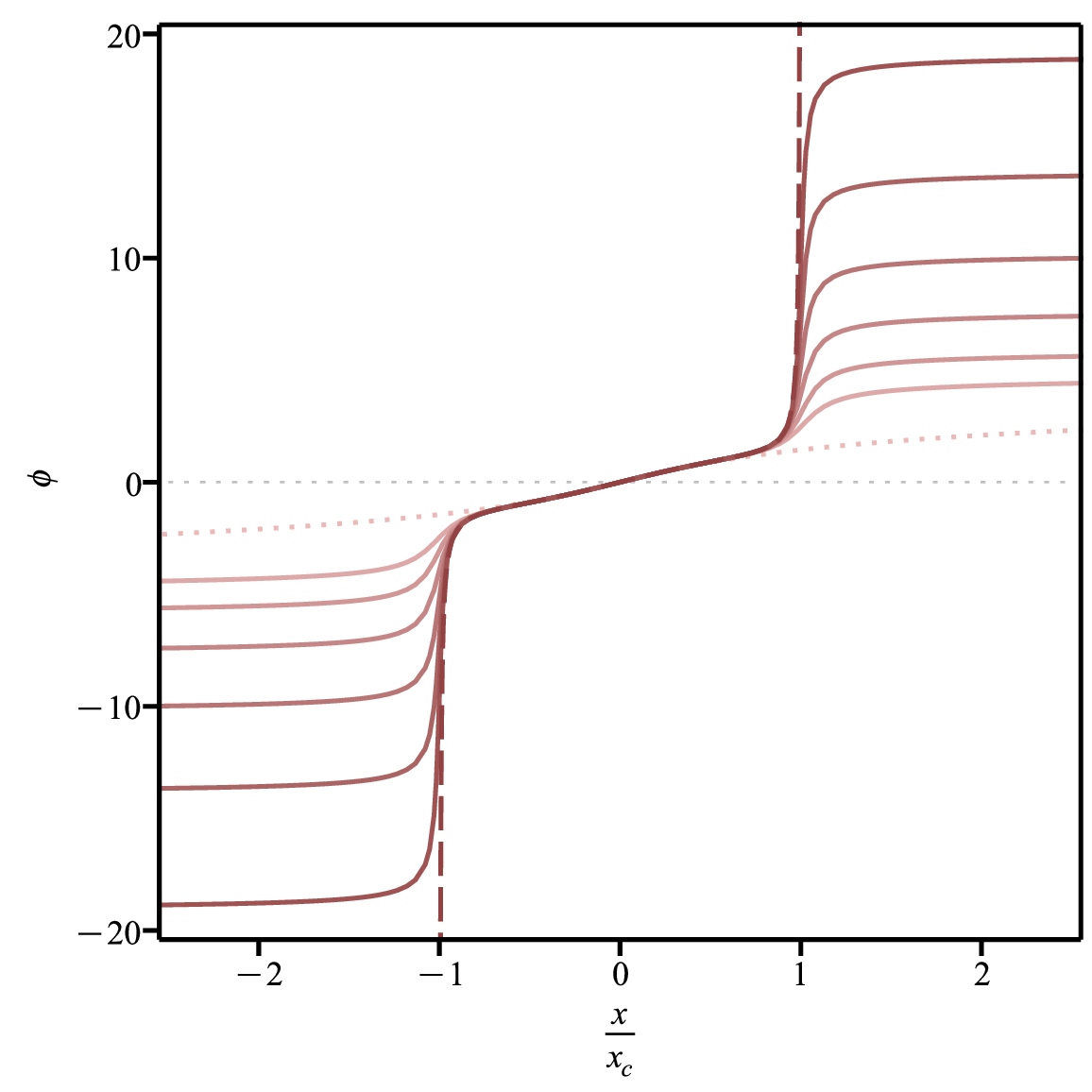}\includegraphics[width=0.333\linewidth]{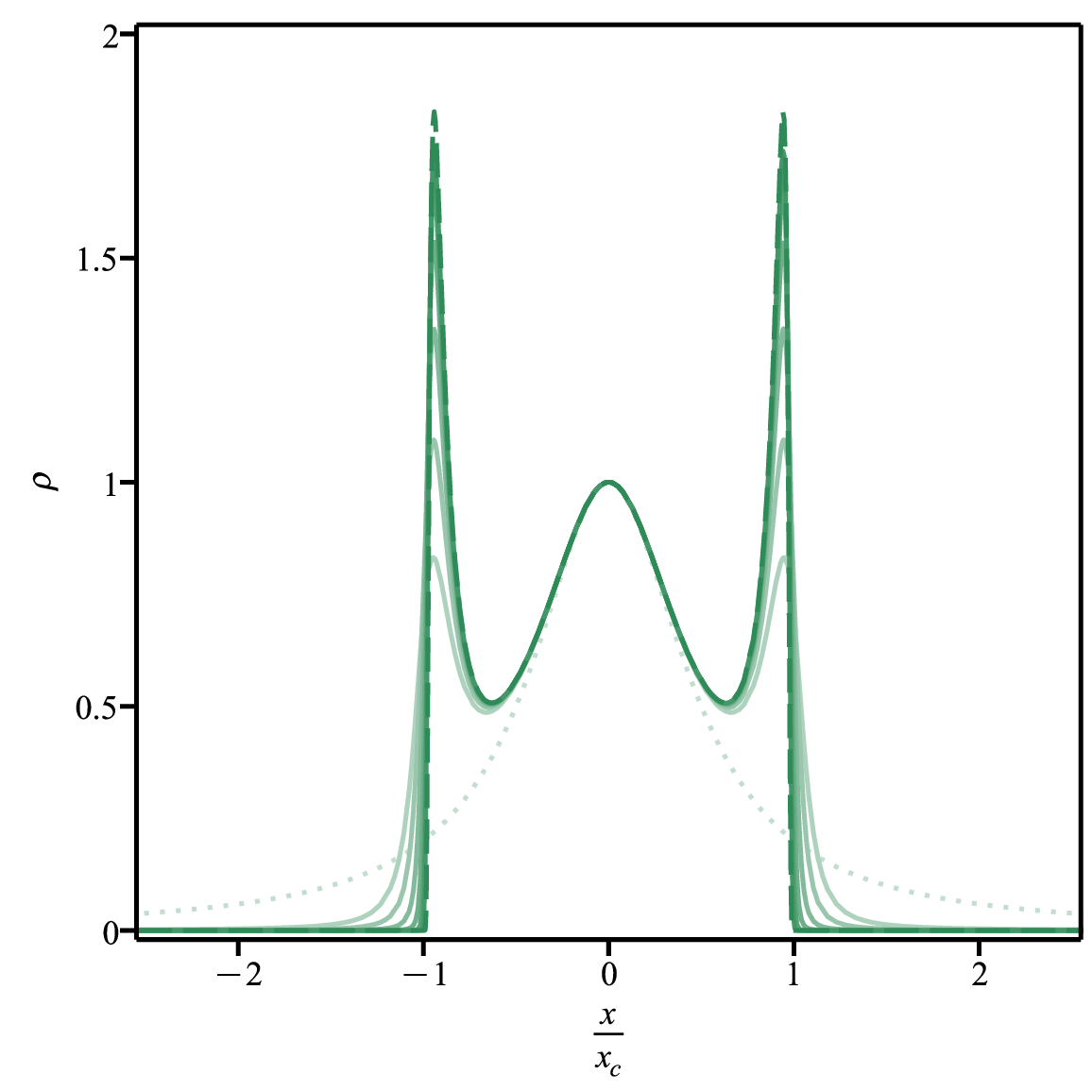}\includegraphics[width=0.333\linewidth]{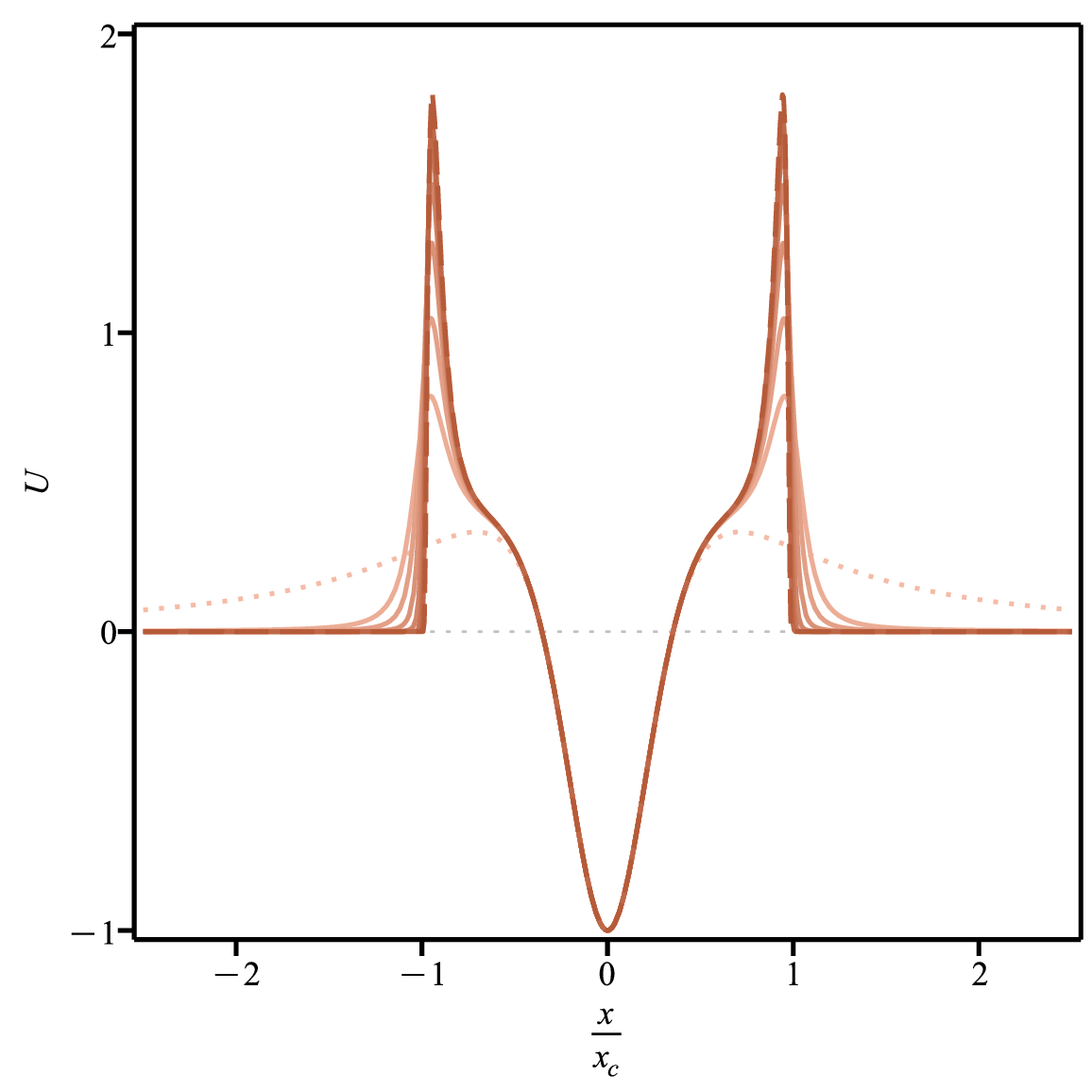}
\caption{The solution $\phi(x)$ of the first-order equation \eqref{fonum} (left) calculated with the condition $\phi(0)=0$, its energy density \eqref{rhonum1} (middle) and the stability potential \eqref{Unum1} (right) for the same line styles and values of $\alpha$ in Fig.~\ref{figsigma1}. The top, middle and bottom panels stand for $x_c=1/2,1$ and $2$.}
\label{fignum1}
\end{figure}

In Fig.~\ref{fignum1}, we display the symmetric solution of \eqref{fonum}, obtained with $\phi(0)=0$, its energy density \eqref{rhonum1} and stability potential \eqref{Unum1} for some values of $x_c$ and $\alpha$. As $\alpha$ gets larger and larger, the tails of the solution get vertically more and more distant from $\phi=0$. For $\alpha\to\infty$, we get a \emph{compact} solution, as expected from the discussion in the last paragraph. In this limit, the energy density and the stability potential also compactify into the interval $x\in[-x_c,x_c]$, vanishing otherwise. Notice that $U(x)$ has a compact-volcano shape, which is also a novel feature and, contrary to what occurs with compact solutions in the impurity-free models \cite{comp2,comp3,k2c}, where $U(x)\to\infty$ outside the compact space and the system admits infinite bound states (without unbound states), the stability potential \eqref{Unum1} is null in the aforementioned interval. Hence, it only supports unbound states and a single bound state, the zero mode, similarly to the impurity-free vacuumless model. This property is quite interesting, as the volcano shape of the stability potential is related to $V(\phi)$, even though the compactification is induced by the impurity.

It is also worth investigating the effects induced by the parameter $x_c$, which controls the size of the compact space. Notice that, as $x_c$ gets larger, the maxima of the impurity \eqref{sigmanum1} gets farther away from the origin. This impacts the energy density and stability potential starkly. Looking at the $\alpha\to\infty$ case, we notice that, as $x_c$ increases, the behavior of the energy density at its core is modified; it is a minimum for $x_c\approx0$ and becomes a maximum for large values of $x_c$. The limit $x_c\to\infty$ breaks the compact support, leading to extended solutions. On the other hand, for $x_c=0$, the impurity gets only a single peak which leads to the singular kink found in the presence of $\sigma^\infty_I(x)$, investigated in Sec.~\ref{secsigmaI}.

We have then seen that the presence of $\sigma_{II}(x)$ is capable of compactifying symmetric solutions. Let us now investigate what occurs with the asymmetric ones. To do so, we consider the condition $\phi(x_0)=0$ with $x_0\neq0$. As we shall see next, the non-null $x_0$ may lead to three distinct types of configurations obeying the first-order equation \eqref{fonum}. We remark that, for finite $\alpha$, the asymptotic behavior of $\phi(x)$ remains \eqref{solvacasy} but, for $x\approx x_0$, we have $\phi(x)\approx (1+\sigma(x_0))(x-x_0)$.  

In the limit $\alpha\to\infty$, the first type of solutions arises for $|x_0|<x_c$. In this situation, for points near the divergences of the impurity, $x\approx\pm x_c$ inside the interval $|x|<x_c$, the solution behaves as $\phi(x)\propto 1/(x_c\mp x)$. To preserve continuity, we suppose that $\phi(x)=\text{sgn}(x)\infty$ outside the limited interval $|x|<x_c$. Therefore, for $|x_0|<x_c$, we get \emph{compact} solutions. The second type of solutions with $\alpha\to\infty$ appears for $x_0>x_c$ or $x_0<-x_c$. Since the latter can be related to the first via the change $x\to-x$, we only consider $x_0>x_c$. In this case, the right tail is not compactified; it falls off obeying \eqref{solvacasy}. The left tail, nonetheless, is compactified. For $x>x_c$ with $x\approx x_c$, the solution behaves as $\phi(x)\propto -1/(x-x_c)$, reaching the vacuum in a limited $x$. To get continuous solutions, we take $\phi(x)\to\infty$ for $x\leq x_c$. This leads to a \emph{half-compact} profile. The third type of solutions in the $\alpha\to\infty$ case emerges for $x_0=\pm x_c$. In this situation, one gets a shifted singular kink, $\phi(x)=\text{sgn}(x-x_0)\,\infty$, with energy density $\rho(x) = \pi\delta(x-x_0)$.

\begin{figure}
\centering
\includegraphics[width=0.333\linewidth]{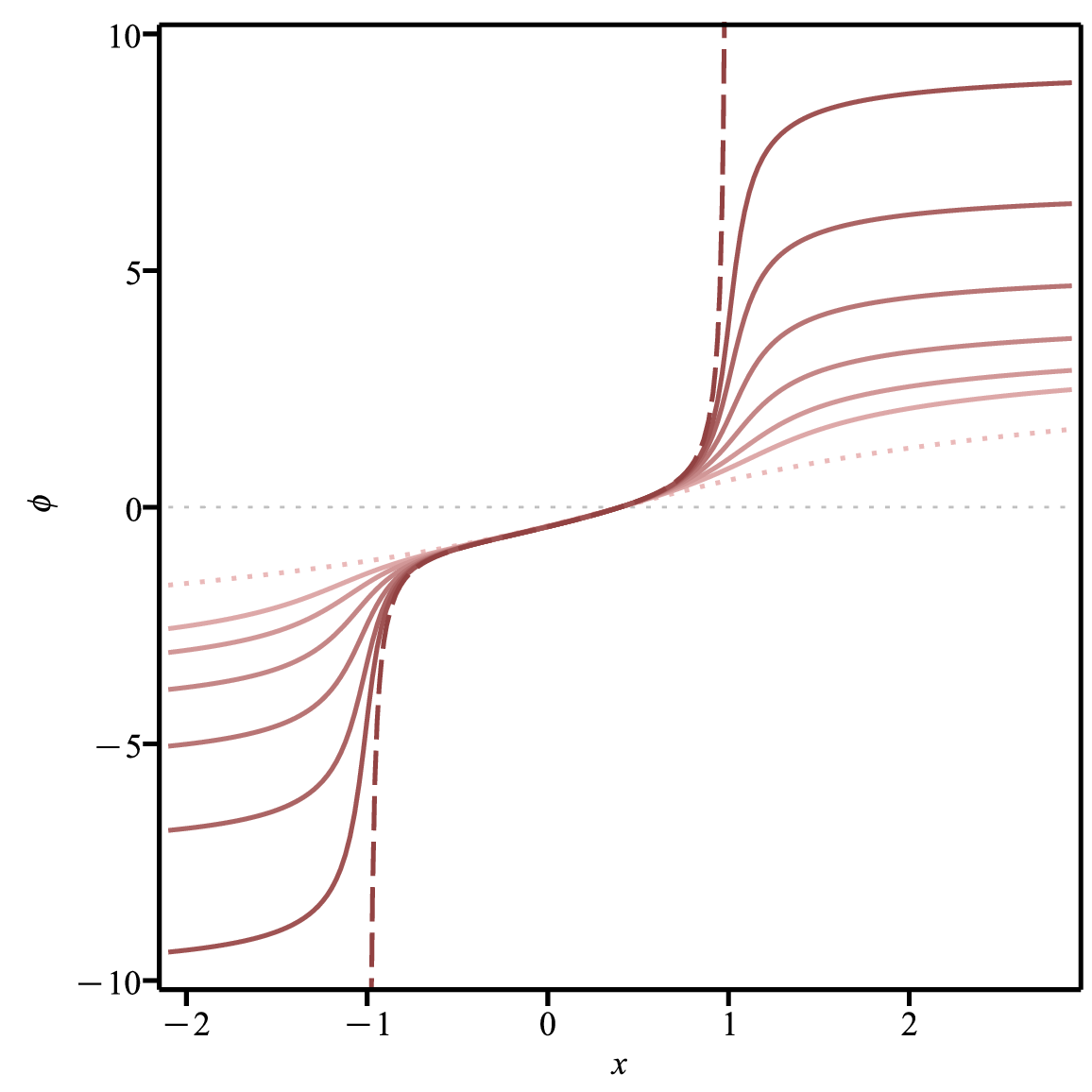}\includegraphics[width=0.333\linewidth]{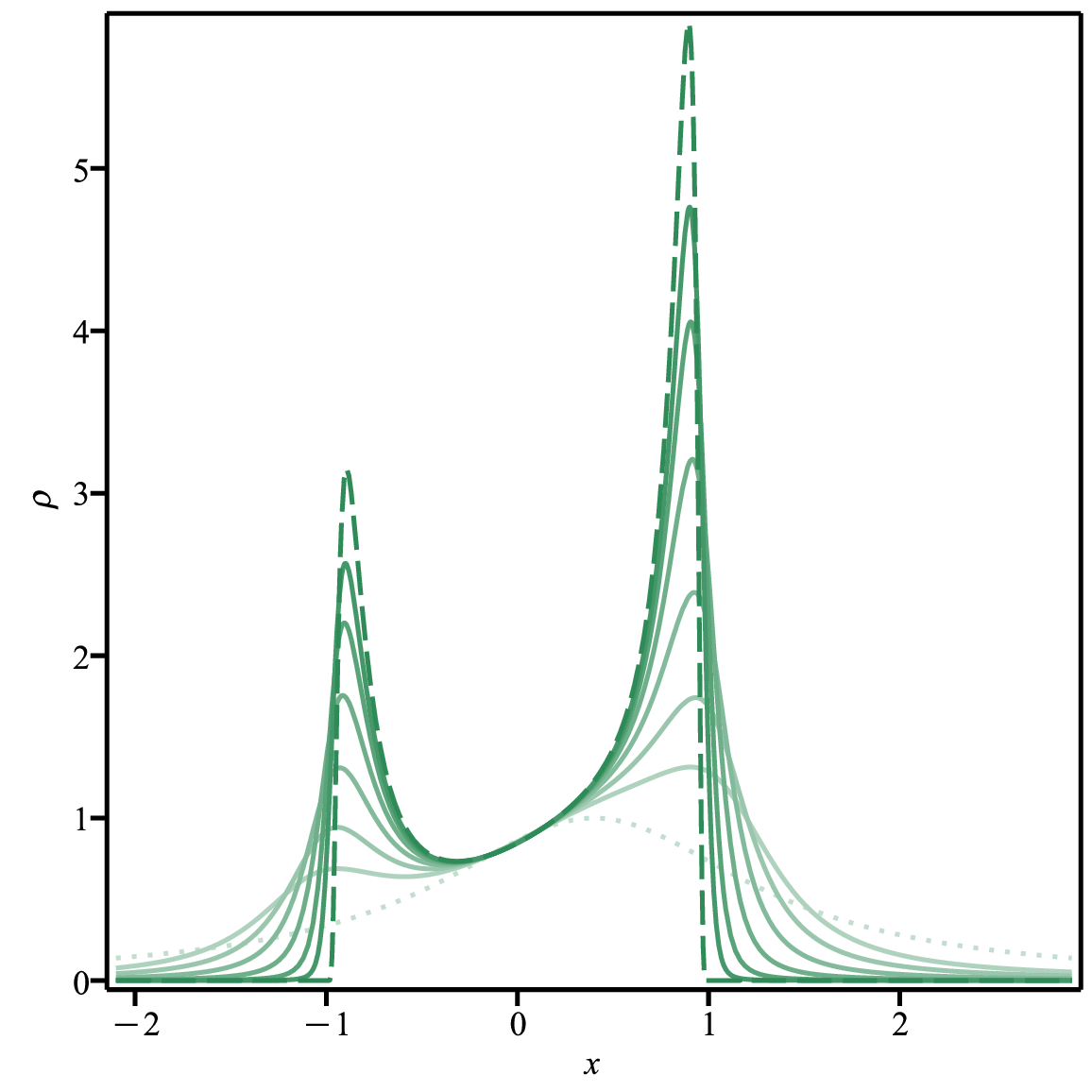}\includegraphics[width=0.333\linewidth]{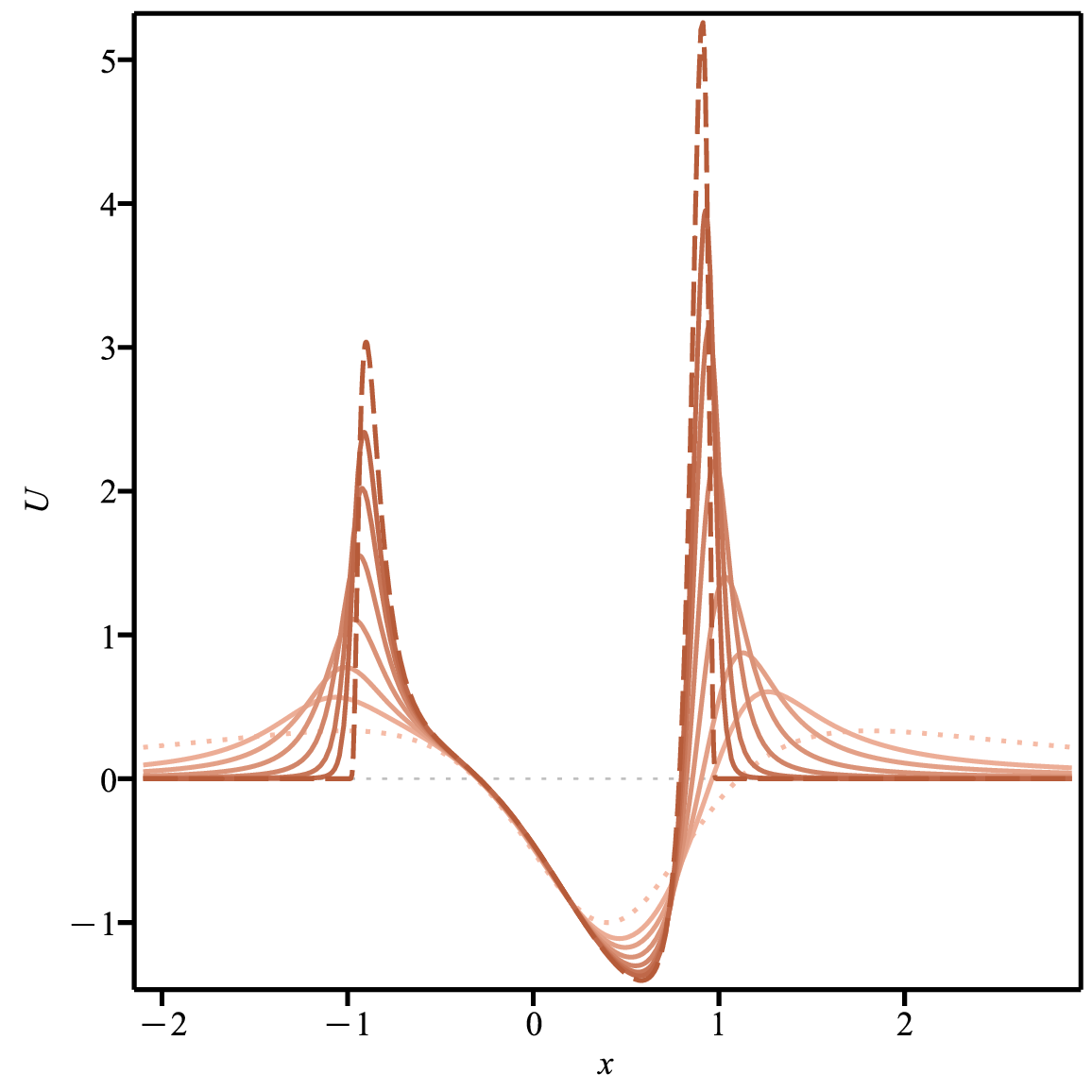}
\includegraphics[width=0.333\linewidth]{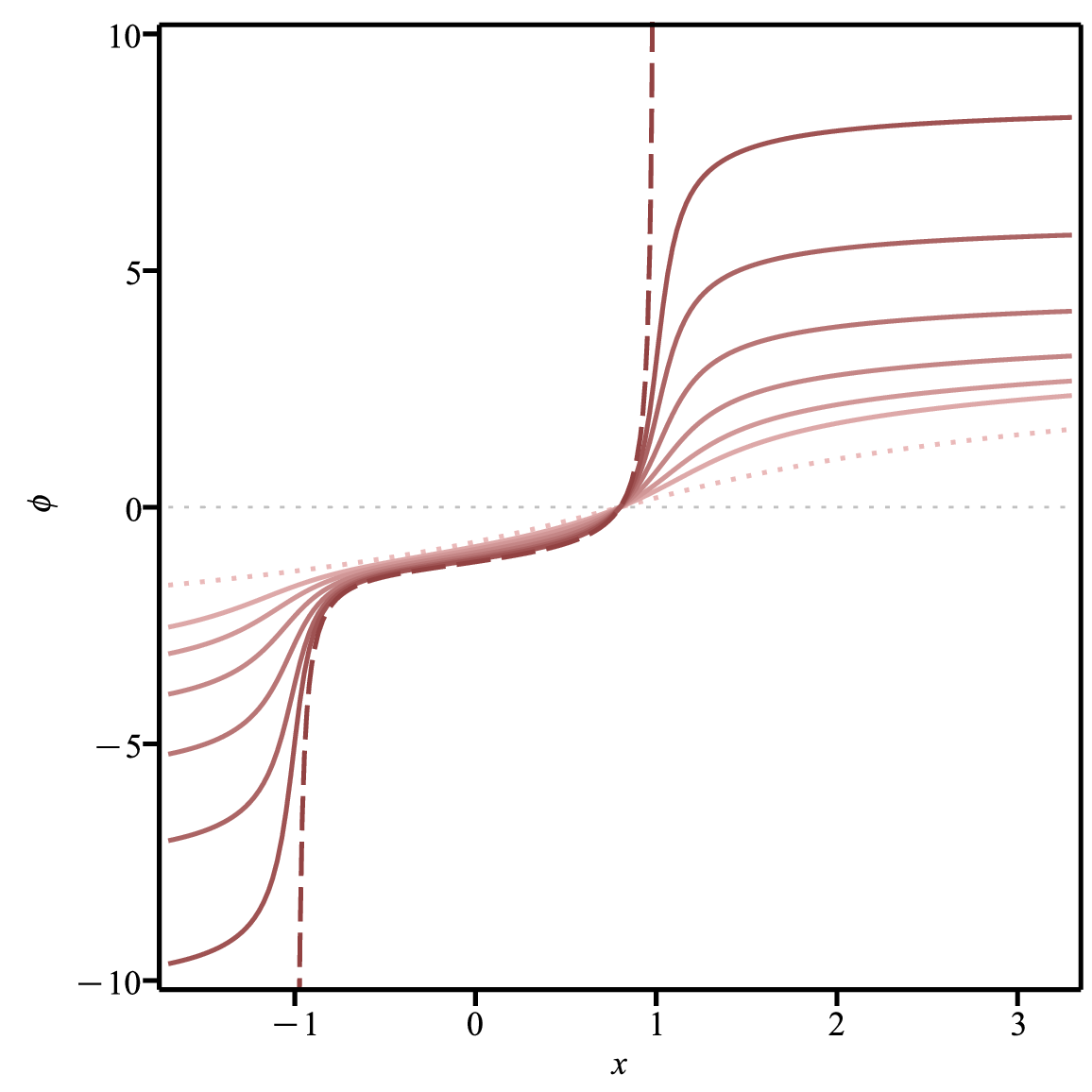}\includegraphics[width=0.333\linewidth]{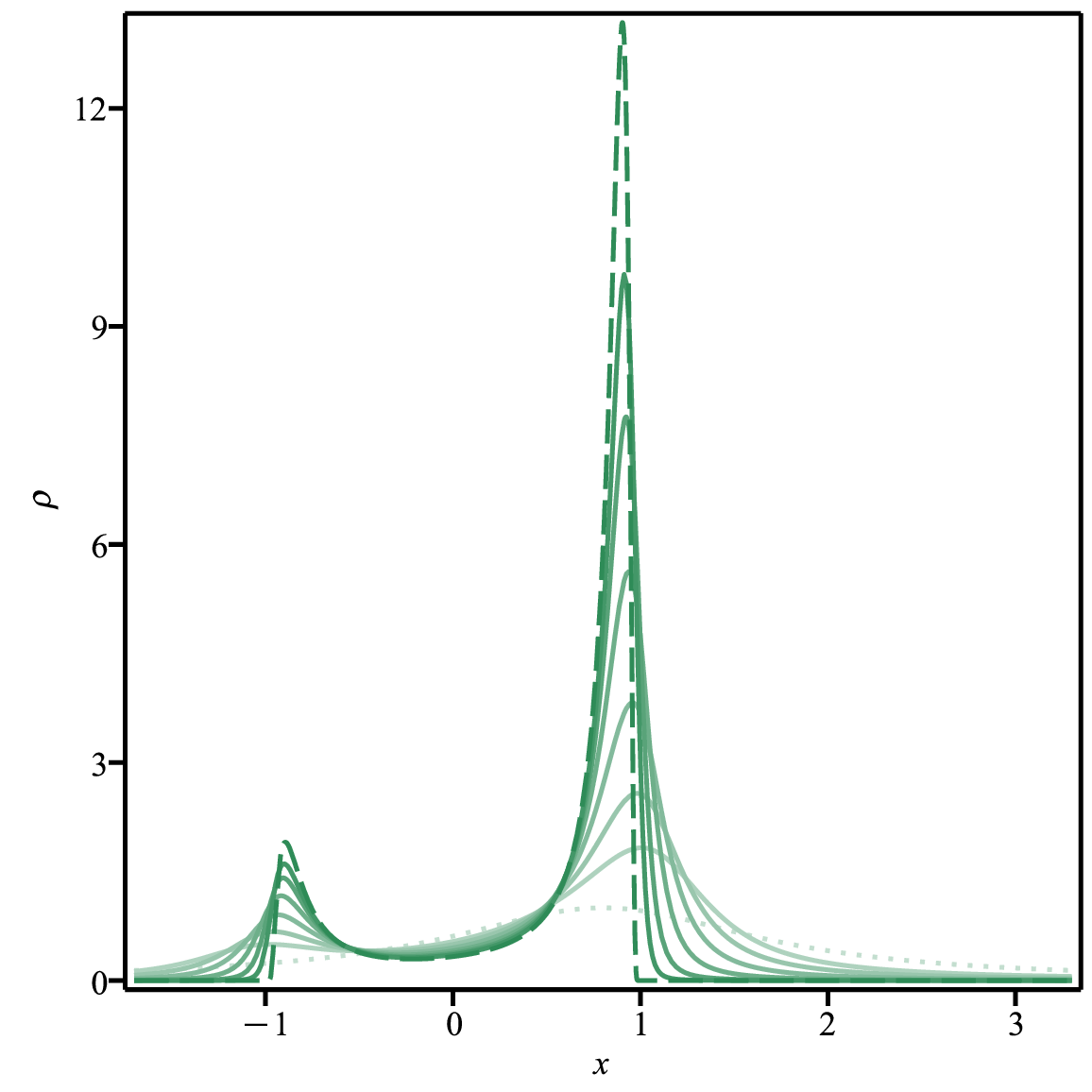}\includegraphics[width=0.333\linewidth]{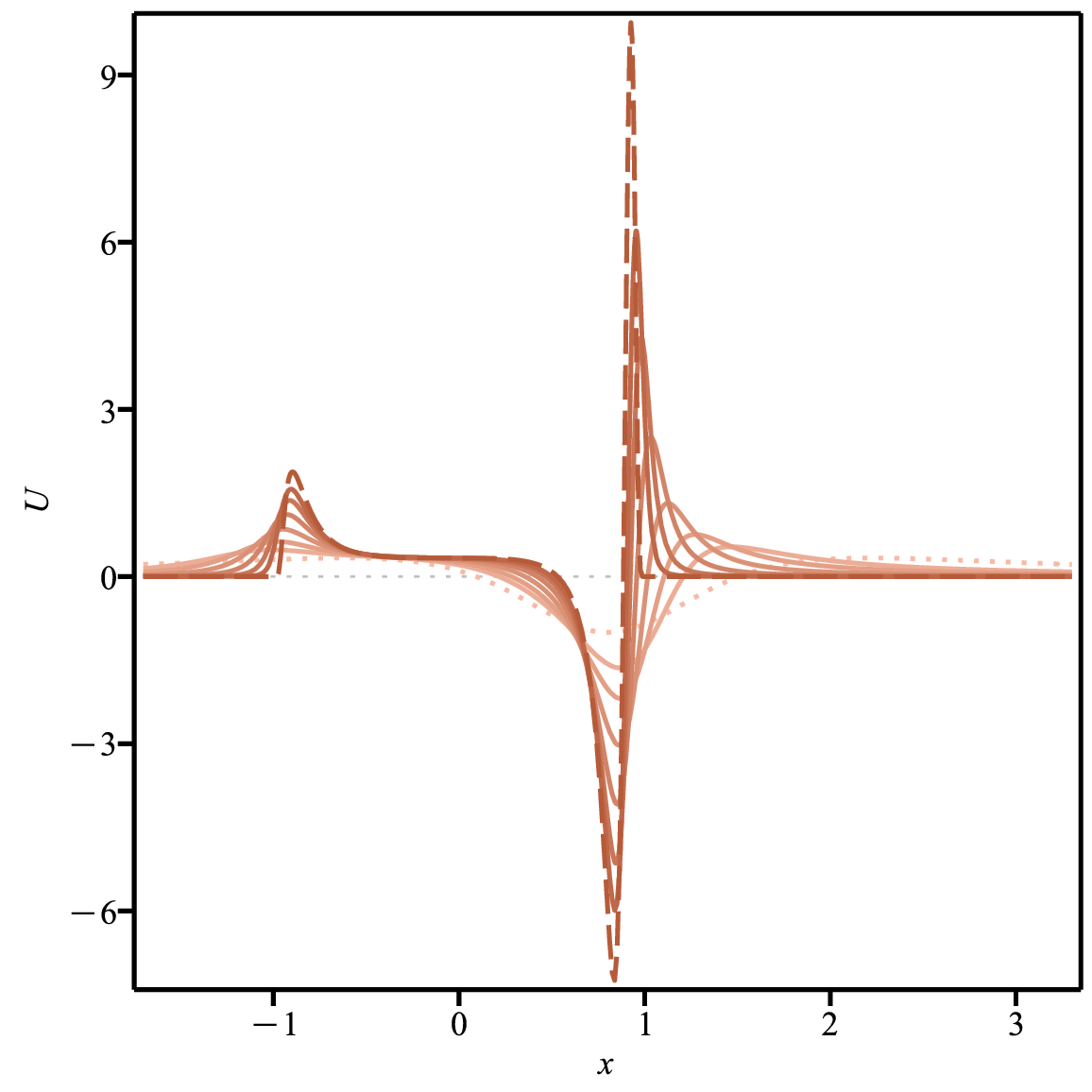}
\includegraphics[width=0.333\linewidth]{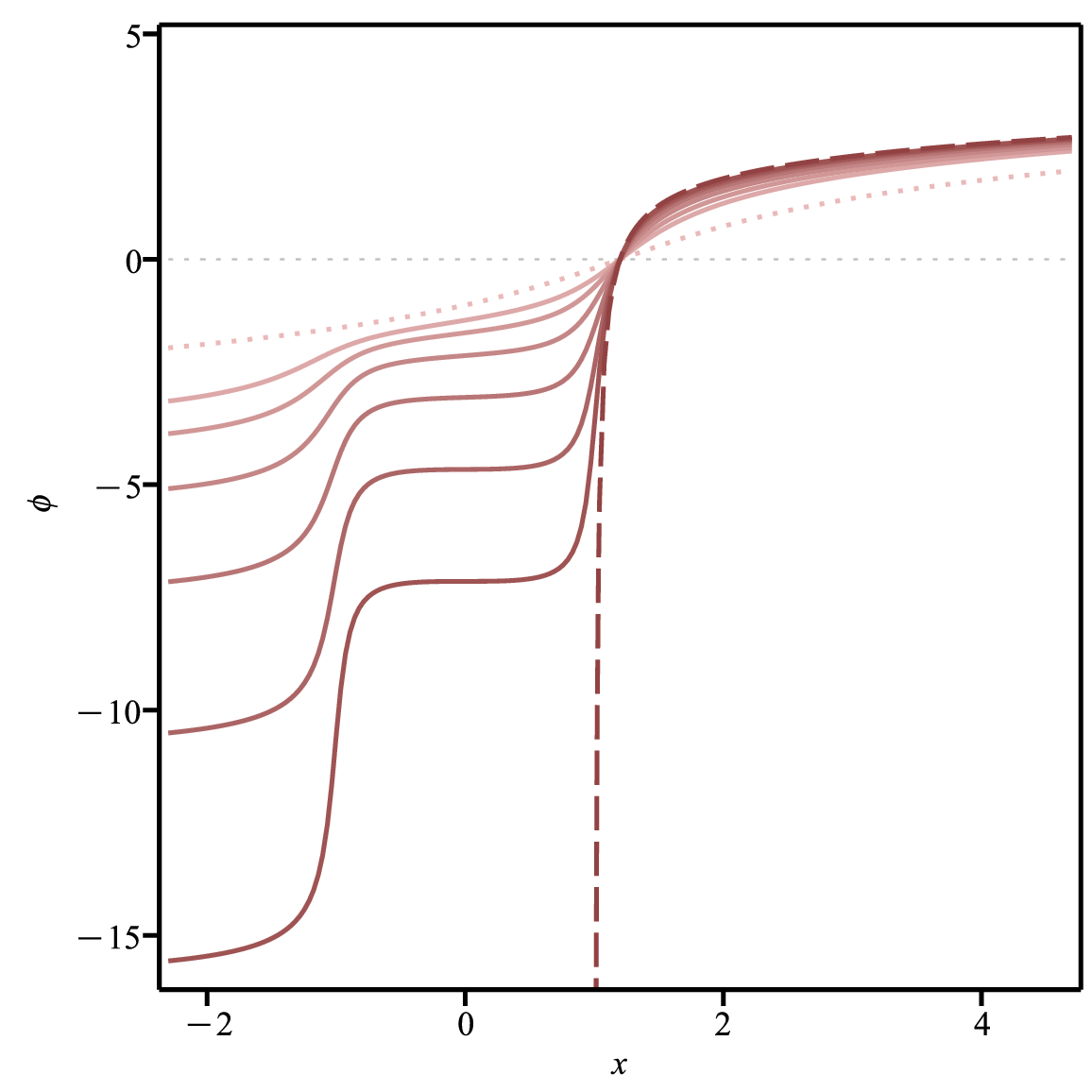}\includegraphics[width=0.333\linewidth]{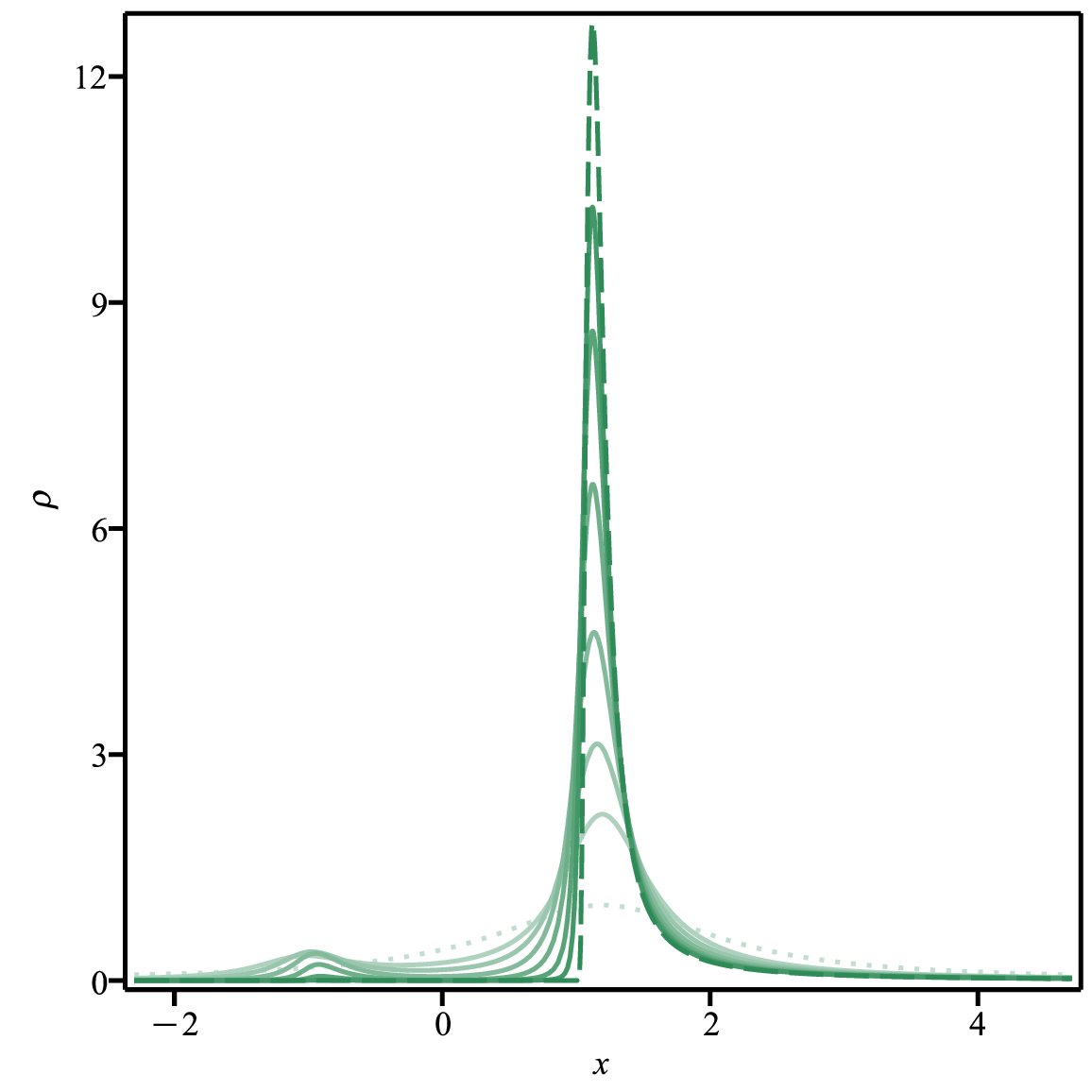}\includegraphics[width=0.333\linewidth]{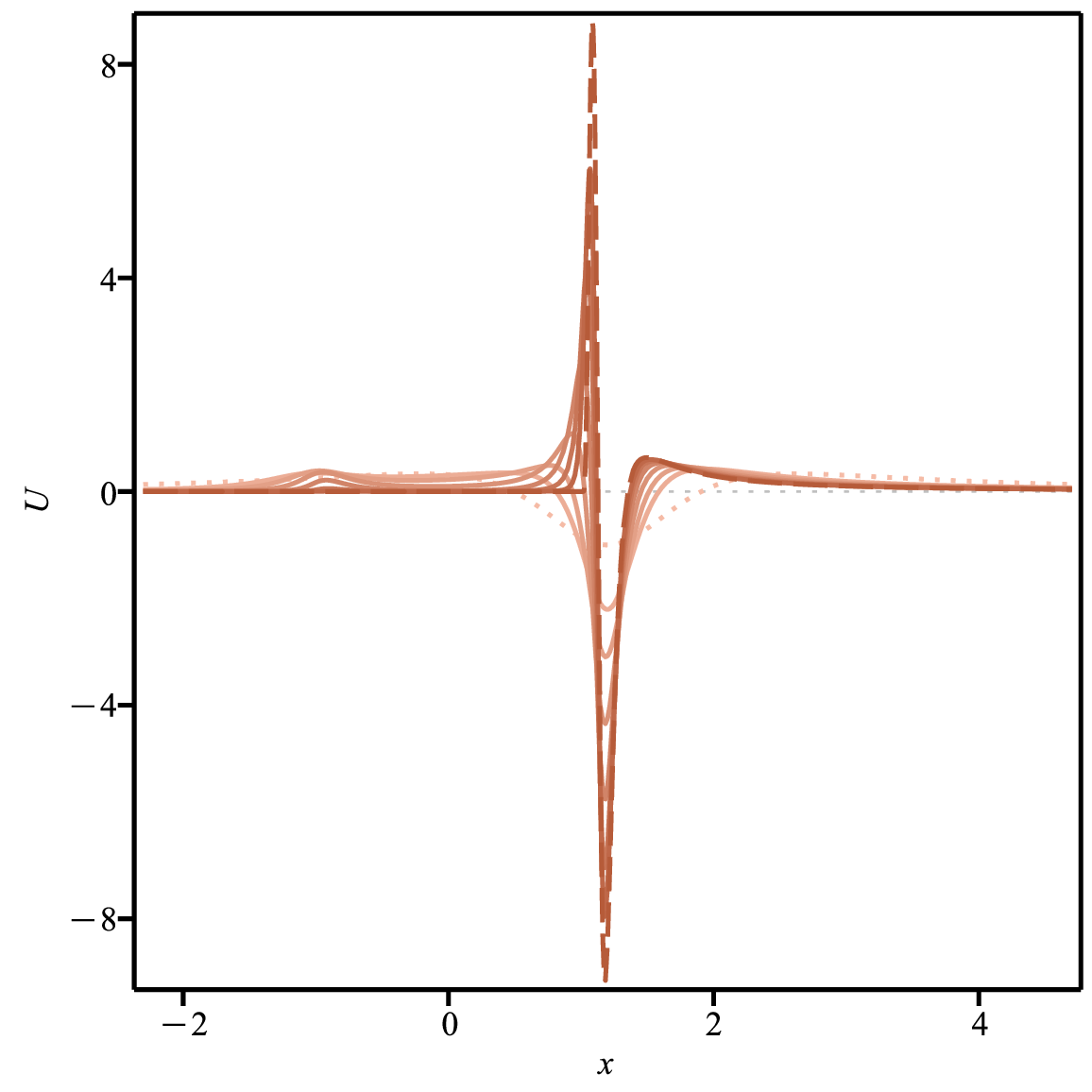}
\includegraphics[width=0.333\linewidth]{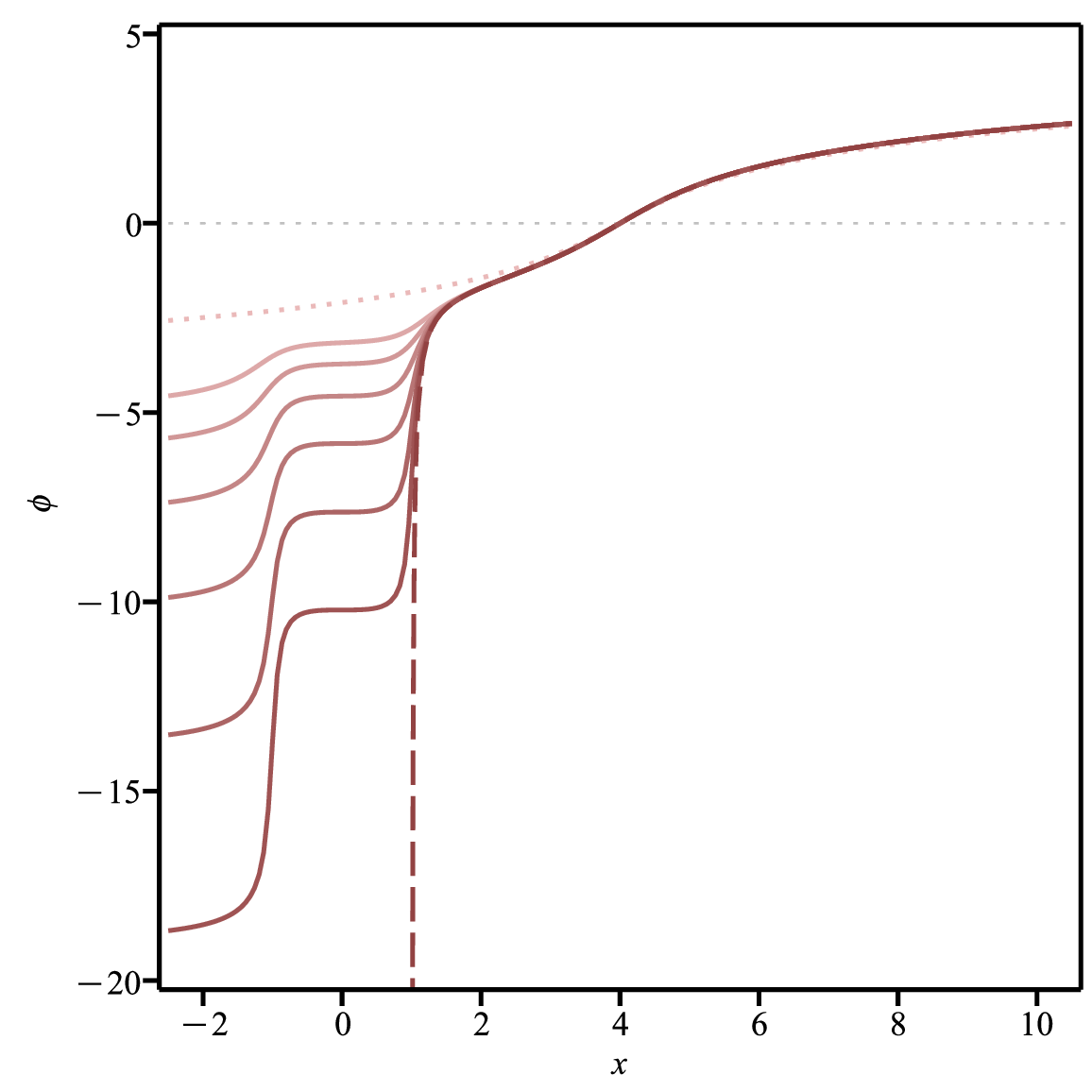}\includegraphics[width=0.333\linewidth]{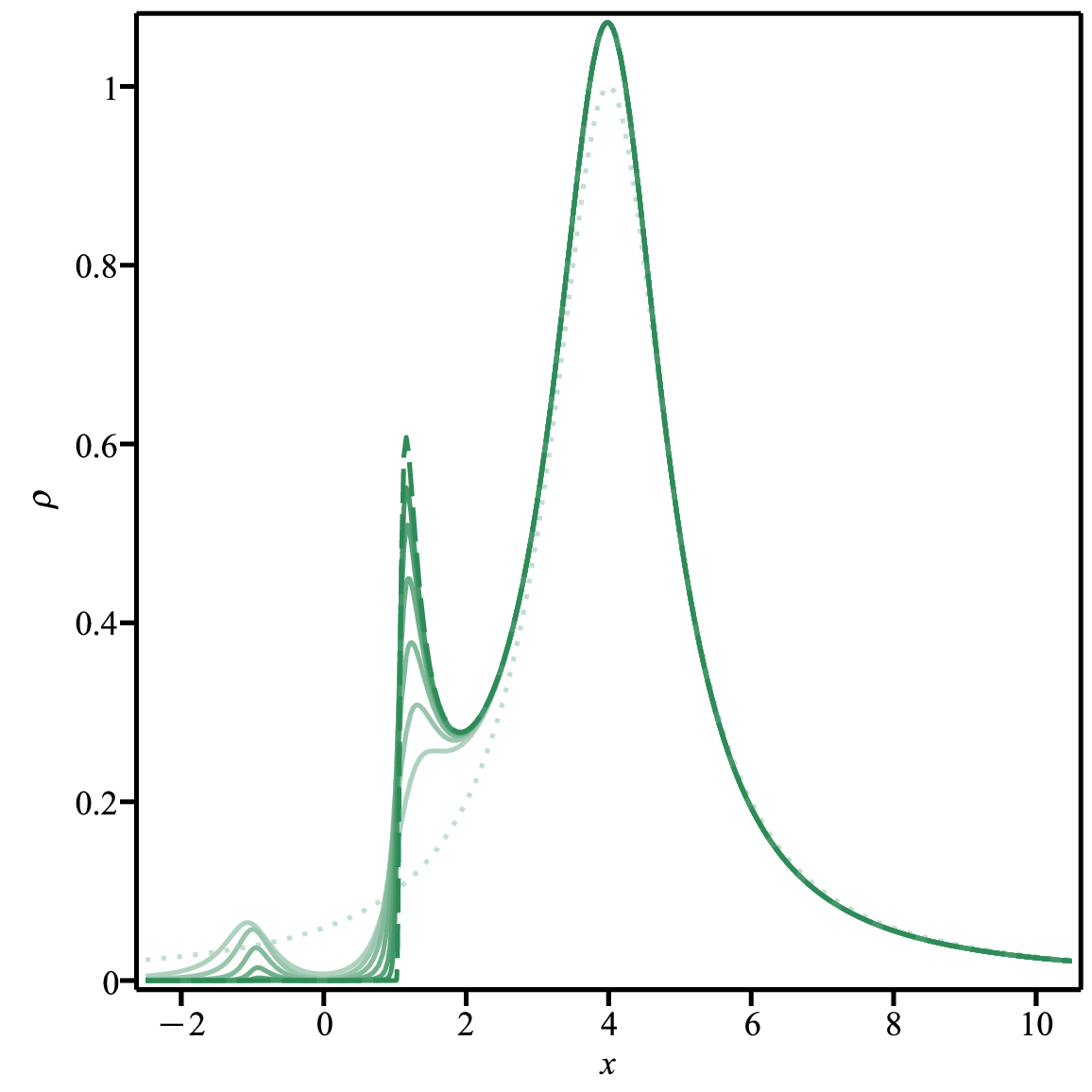}\includegraphics[width=0.333\linewidth]{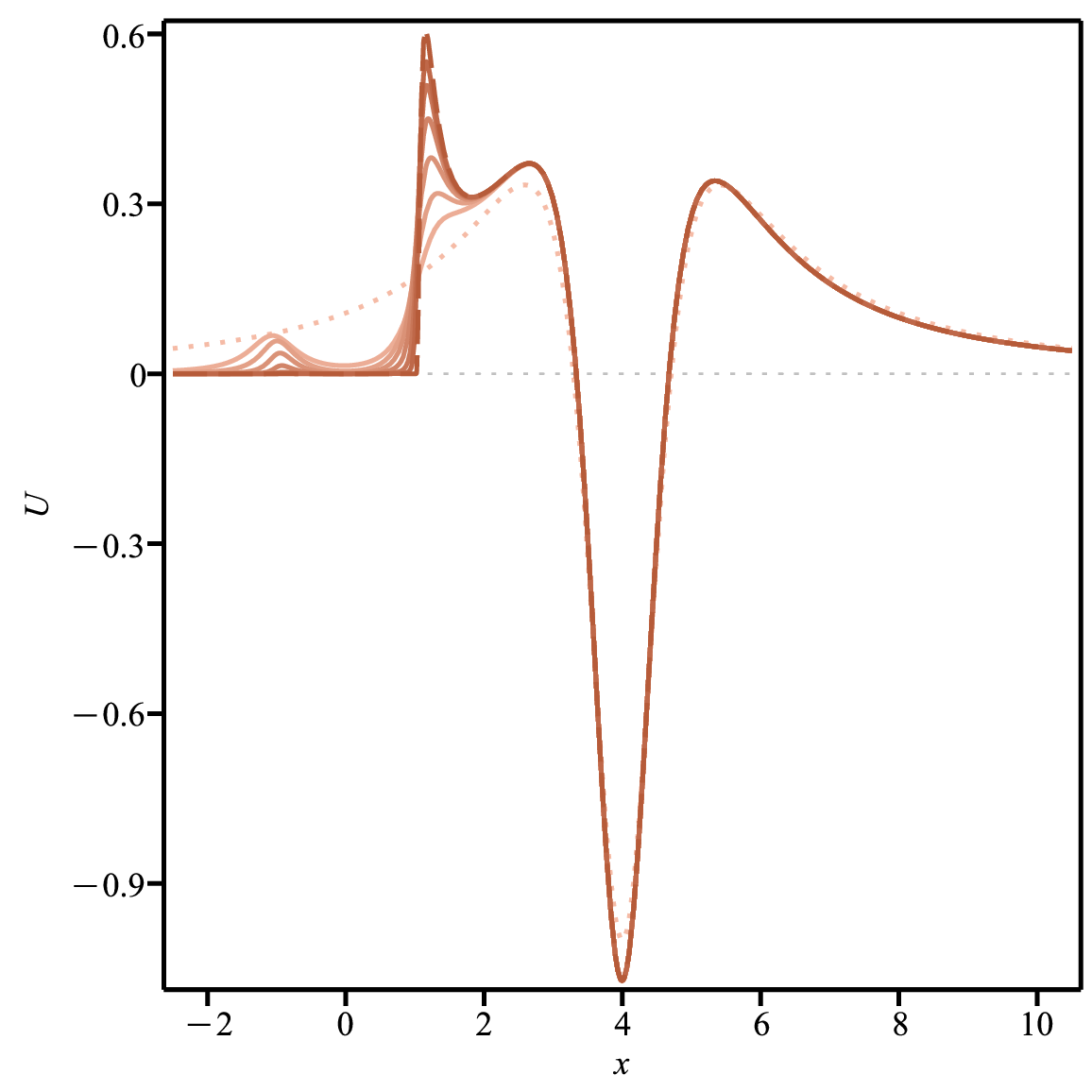}
\caption{The solution $\phi(x)$ of the first-order equation \eqref{fonum} (left) calculated with the condition $\phi(x_0)=0$, its energy density \eqref{rhonum1} (middle) and the stability potential \eqref{Unum1} (right) for the same line styles and values of the parameters $x_c$ and $\alpha$ in Fig.~\ref{figsigma1}. The top, middle-top, middle-bottom and bottom rows are for $x_0=0.4,0.8,1.2$ and $4$, respectively.}
\label{fignum2}
\end{figure}

To illustrate the behavior of the structure for $x_0\neq0$, we plot the solution of Eq.~\eqref{fonum}, the energy density \eqref{rhonum1} and the stability potential \eqref{Unum1} in Fig.~\ref{fignum2} for $x_c=1$ and several values of $\alpha$ and $x_0$. As expected, $x_0\neq0$ makes the aforementioned physical quantities become asymmetric. For $|x_0|<1$, we see that the parameter $\alpha$ compactifies both tails of the solution. However, for $x_0>1$, the parameter $\alpha$ only compactifies the tail which would cross the divergence that emerges in the impurity for limit $\alpha\to\infty$, giving rise to a half-compact solution. 

\section{Outlook}\label{secoutlook}
In this manuscript, we have investigated how to compactify the kink solutions that emerge in vacuumless systems with impurities. Working with the Lagrangian density \eqref{lmodel}, we consider specific constraints for the functions associated to the self-interactions of the scalar field, which is required to be static, in order to obtain first-order equations which lead to minimum-energy configurations compatible with the equations of motion. We also inspect the linear stability, to verify how the static solution behaves around small fluctuations. This leads to a Sch\"odinger-like equation with zero mode depending on the derivative of the solution and the impurity.

We have briefly reviewed the basic properties of the impurity-free ($\sigma=0$) vacuumless system \cite{vacuumless1,vacuumless2} described by the auxiliary function \eqref{wvac} within the model \eqref{lmodel} to show that the vacuumless kink diverges logarithmically, ranging from $-\infty$ to $\infty$ as $x$ goes from $-\infty$ to $\infty$. Even so, its energy density is localized, vanishing asymptotically, and the energy is finite. By considering the model \eqref{lmodel}, we have used the Cauchy-Schwarz inequality to show that, without impurities, one \emph{cannot} compactify the vacuumless kink in such model because it would lead to infinite energy.

The impurity $\sigma(x)$ enters the first-order equation additively, so it may be used to increase or decrease the slope of the solution at specific points. We then presented two impurity functions, \eqref{sigmaI} and \eqref{sigmanum1}, which are able to induce compactification of the solution via a parameter $\alpha$, exploring their effects on the energy density and stability potential. An interesting new result indicates that the presence of impurities significantly changes the physical characteristics of the solutions.

To illustrate the main results, we have studied two different situations. First, we have considered $\sigma_I(x)$, which presents a single maximum at the origin whose height is given by $\alpha$. In the limit $\alpha\to\infty$, the maximum becomes a divergence. In this situation, we have shown that solutions obeying $\phi(x_0)=0$ become half-compact for $x_0\neq0$ and, remarkably, the symmetric solution ($x_0=0$) has the form of the singular kink \cite{asen1,asen2,trilogia1}.
The presence of $\sigma_{II}(x)$ was also investigated. In this scenario, the impurity engenders two symmetric maxima around the origin. For $\alpha\to\infty$, these points become divergences located at $x=\pm x_c$. This case is richer. We have shown that the solutions may become half-compact, compact or singular, depending on the value of $x_0$ in the condition $\phi(x_0)=0$ used to solve the first-order equation.

As perspectives, one can investigate impurities with other properties to verify how they modify kink solutions in vacuumless systems described by \eqref{wvac}. For example, a parameter $s$ can be included that generalizes $\sigma_{II}(x)$ to
\be\label{sigmas}
\sigma(x) = \frac{\alpha\,x^2(1-2sx_c^2+sx^2)}{1+\alpha\left(x_c^2-x^2\right)^2},
\ee
with $s=[0,1/(2x_c^2)]$. Notice that $\sigma_{II}(x)$ is recovered for $s=0$. This impurity has two maxima, similarly to $\sigma_{II}$, but its asymptotic behavior is $\sigma(x)\approx s + 1/x^2$, so $\sigma(x)\to s$ for $x\to\pm\infty$. Therefore, it introduces changes in the tail of the kink. Indeed, for finite $\alpha$ and $s\neq0$, one can show that $\phi(x)\propto x$ asymptotically, which is a straight line instead the logarithmic function \eqref{solvacasy}. Interestingly, for $\alpha\to\infty$, the symmetric solution ($\phi(0)=0$) obeys $\phi(x)\propto 1/(x_c\mp x)$ for $x\approx x_c$ with $|x|<x_c$, giving rise a compact profile.

Another possibility for future investigation concerns analytical solutions. One may consider, for example, a field configuration with the form
\be\label{solanac}
\begin{aligned}
\phi(x) &= \frac{x}{1+c^2} +\frac{c^2}{2\left(1+c^2\right)\sqrt{1-c^2}}\\
    &\times\left(\arctan\left(\frac{x+c}{\sqrt{1-c^2}}\right)+\arctan\left(\frac{x-c}{\sqrt{1-c^2}}\right)\right),
\end{aligned}
\ee
where $c\in[0,1)$ and construct an impurity compatible with it for the auxiliary function \eqref{wvac}. The asymptotic behavior is $\phi(x)\propto x$, similarly to the one arising with \eqref{sigmas}. In the above expression, the case $c=0$ simplifies to $\phi(x)=x$, which connects $-\infty$ to $\infty$ as $x$ spans from $-\infty$ to $\infty$. The limit $c\to 1$ leads to the compact solution
\be
\phi(x) = \begin{cases}
    \cfrac{x}{2} +\cfrac{x}{2(1-x^2)},&|x|<1\\
    \text{sgn}(x)\,\infty,&|x|
    \geq1.
\end{cases}
\ee
It is worth noting that the above expression and \eqref{solanac} produce odd symmetry, obeying the initial condition $\phi(0)=0$. One may also try to obtain half-compact solutions described by analytical functions by considering condition $\phi(x_0)=0$ with $x_0\neq0$ in the first-order equation.

It is also of interest to recall that, if one thinks of the impurity as an isolated inhomogeneity, it would be described by the Dirac distribution. This type of impurity was previously studied long ago, for example, in Ref. \cite{imp3}. Moreover, one knows that the Gaussian, Lorentzian, and hyperbolic secant can be used as good representations of the Dirac distribution, so one may also consider the Gaussian and hyperbolic secant to describe localized generalizations of an isolated inhomogeneity, which will be further discussed in a separate work.
Another possibility is to study vacuumless gauge vortices in two spatial dimensions in the presence of the local $U(1)$ symmetry, and vacuumless gauge monopoles in three spatial dimensions in the presence of the local $SU(2)$ symmetry. Some investigations appeared previously in Refs. \cite{vacuumless1,vlvortex} in the absence of impurities, and we think it is worth adding impurities to search for different behavior. For example, 
some results previously obtained in Ref. \cite{vlvortex} show that vortices in models with vacuumless scalar fields present a large tail that extends far away from the origin, but with the energy density localized and the total energy finite. Moreover, the magnetic flux is well defined and still works as a topological invariant. These properties suggest that it is of current interest to investigate the model in the presence of impurities. The study of solutions in vacuumless systems doped with impurities may also be considered in curved spacetime \cite{impcurvo}. In this scenario, the components of the metric enter the first-order equation, so other conditions may be needed for the potential and/or impurity, bringing new possibilities for modeling the system.

\acknowledgments{We acknowledge financial support from the Brazilian agencies Conselho Nacional de Desenvolvimento Científico e Tecnológico (CNPq), grants Nos. 402830/2023-7 (DB, MAM and RM),  303469/2019-6 (DB), 306151/2022-7 (MAM) and 304344/2025-7 (RM), and Paraiba State Research Foundation (FAPESQ-PB) grant No. 2783/2023 (IA).}



\end{document}